\documentclass[12pt]{article}
\usepackage{latexsym}
\usepackage{amsmath}
\usepackage{epsfig,graphics}

\newcommand{\be}{\begin{equation}}
\newcommand{\ee}{\end{equation}}

\newlength{\figsize}
\begin{document}

\begin{titlepage}

\vspace*{0.7in}
 
\begin{center}
{\large\bf On the spectrum of closed k=2 flux tubes \\
in D=2+1 SU(N) gauge theories\\}
\vspace*{1.0in}
{Andreas Athenodorou$^{a}$, Barak Bringoltz$^{b}$  and Michael Teper$^{a}$\\
\vspace*{.2in}
$^{a}$Rudolf Peierls Centre for Theoretical Physics, University of Oxford,\\
1 Keble Road, Oxford OX1 3NP, UK\\
\vspace*{.1in}
$^{b}$Department of Physics, University of Washington, Seattle,
WA 98195-1560, USA\\
}
\end{center}

\vspace*{0.55in}

\begin{center}
{\bf Abstract}
\end{center}

We calculate the energy spectrum of a $k=2$ flux tube that is
closed around a spatial torus, as a function of its length $l$. We do 
so for SU(4) and SU(5) gauge theories in 2 space dimensions.
We find that to a very good approximation the eigenstates belong
to the irreducible representations of the SU($N$) group rather than 
just to its center, $Z_N$. We obtain convincing evidence that
the low-lying states are, for $l$ not too small, very close to those
of the Nambu-Goto free string theory (in flat space-time). The 
correction terms appear to be typically of $O(1)$ in appropriate 
units, much as one would expect if the bosonic string model were 
an effective string theory for the dynamics of these flux tubes. 
This is in marked contrast to the case of fundamental flux tubes 
where such corrections have been found to be unnaturally small. 
Moreover we 
find that these corrections appear to be particularly small when
the `phonons' along the string have the same momentum, and 
large when their momentum is opposite. This provides
information about the detailed nature of the interactions 
in the effective string theory. We have searched for, but not
found, extra states that would arise from the excitation of
the massive modes presumably associated with the non-trivial  
structure of the flux tube.

\end{titlepage}

\setcounter{page}{1}
\newpage
\pagestyle{plain}

\section{Introduction}
\label{section_intro}

Consider a $D=2+1$ SU($N$) gauge theory on a three-torus, and
a loop of confining flux that winds once around a 
spatial torus of length $l$. If this flux is in the fundamental
representation then one discovers 
\cite{gs_k1,es_k1},
that the low-lying energy spectrum of such a loop is the
same as that of a corresponding closed string in the Nambu-Goto 
free string theory (in flat $D=2+1$ space-time) up to very small
corrections, and that this is so even when $l$ is not much larger
than the critical length at which the flux loop dissolves. 
This type of calculation is intended to provide detailed information
about the effective string theory describing the gauge theory
and should be useful for theoretical approaches from the
string side.  

In the usual field-theoretic picture of the confining string 
as some non-Abelian version of the flux tube in a (dual) superconductor, 
with some complicated internal spatial structure which should
provide a complex set of internal modes to excite,
it is not easy to understand how a closed flux tube that
is not much longer than it is wide, can display the phonon-like
excitations of a thin string to such a good accuracy. It is therefore
tempting to see this as evidence for some underlying duality
where the starting point in describing a flux tube is always a 
thin string, irrespective of its length, just as one has in the 
usual gauge/gravity duality. (See
\cite{ads_rev}
for some recent reviews.) Even in the latter framework, however,
one does expect states in the spectrum that are additional
to those of the free bosonic string and which arise, for example,
from the non-trivial background metric near the `horizon' which leads
to linear confinement. Finding such extra states 
would provide a useful clue about the nature of the hypothetical
dual theory.
So far no such `extra' states have been identified in our
lattice calculations
\cite{es_k1}
but this may be because the basis of operators we use in our
variational calculation has little overlap onto any such states
(as discussed below).
This is one reason why it is useful to explore the spectrum of
the quite different flux tubes that we study in this paper.

In this paper we explore what happens to the spectrum of 
a winding flux tube when the flux is in a higher representation
than the fundamental. In particular for $N>3$ there exist new, completely
stable flux tubes called $k$-strings. A $k$-string is a flux tube
that emanates from a source which transforms as $\psi \to z^k \psi$
under a global gauge transformation belonging to the center of
the group:  $z\in Z_N$.
One may think of this source as being a composite of $k$ fundamental
sources, together with any number of adjoint sources which do not feel
the centre. (It is clear that $k$
and $k^\prime=N-k$ are the `same' and only $k\le N/2$ is interesting.)
Since $k$ is unchanged by screening with gluons, the lightest
flux tube in this sector will be stable. That is what one usually
calls a $k$-string. (Sometimes we shall refer more loosely to
all such flux tubes, stable and unstable, as $k$-strings.)
While such a stable flux tube is of especial interest, excitations 
thereof, including flux tubes in other representations, will
become stable at larger $N$ and are also of interest.

Given that a single fundamental flux loop is so well described by a
free bosonic string theory, one might have expected that a $k$-string
would simply look like some linear combination of free fundamental ($k=1$) 
strings. So, for example, a $k=2$ string might look like a combination 
of a $k=1$ string that winds twice around the torus, and a state
consisting of two $k=1$ strings, each winding once around the torus.
The free string theory would, in this way, predict a simple relation
between the string tensions: $\sigma_k = k \sigma_f$. However it has
been known for some time 
\cite{kstrings_old3}
that the lightest $k$-string is in fact strongly bound 
(for $k/N$ not too small) and only acquires the 
free-string value at $N=\infty$ (as it must, at fixed $k$).
Given that the binding is strong, e.g. in SU(4) one
finds  $\sigma_{k=2} \simeq 1.34 \sigma_f \ll 2 \sigma_f$, one can 
ask to what extent such a flux tube also behaves like a free bosonic
string and what the corrections to such a behaviour tell us
about the dynamics. 
This binding cannot be readily incorporated into the usual 
analytic frameworks
\cite{LSW,PS}
since it presumably involves the exchange of 
small contractible loops of string (glueballs) that are not under 
analytic control. Thus such flux tubes are of particular interest.

The ground state energy of a closed $k=2$ flux
tube of length $l$ has been calculated in 
\cite{gs_k},
where it was found to be accurately described by the
free Nambu-Goto string, with string tension $\sigma=\sigma_{k=2}$,
and with corrections (in inverse powers of $1/\sigma l^2$) that have
coefficients that are $O(1)$ in natural units. This is what one
would expect if the $k=2$ flux tube were to have an effective bosonic 
string description at large distances, and it provides a contrast with
the fundamental flux tube where the corrections to the
free string description are very much smaller
\cite{gs_k1,es_k1}. 
In practice, however, the ground 
state provides only a limited insight into the nature 
of the string spectrum and so, just as for the fundamental string, 
it is important to perform accurate calculations of at least a 
few excited states. This is difficult in a lattice calculation,
where one can only hope to obtain accurate energies for those 
states which have a very high overlap onto the basis that one is 
using. For the fundamental flux loop we attacked this problem, with 
apparent success, by using a very large basis of operators
\cite{es_k1}.
Here we shall do the same for the $k=2$ spectrum, using the basis 
described in Section \ref{section_methods} and in Appendix A. 
Since we have seen from earlier calculations of this type 
\cite{gs_k1,es_k1,gs_k,glued3}
that lattice spacing corrections are small, we shall not 
perform a continuum extrapolation, but will simply choose
to perform our calculations at fixed, small values
of the lattice  spacing $a$. We shall perform calculations in
SU(4) and SU(5) at $\beta=50$ and $\beta=80$ respectively.

In the next section we briefly describe the standard aspects
of our lattice calculation and then, in some more detail, our
choice of operators for winding flux tubes. We follow that up
with a summary of the corresponding free string Nambu-Goto spectrum,
since that will play an important part in the interpretation
of our results. We then describe our results for the spectrum
and find that we can interpret at least some of the states
within the Nambu-Goto picture, simply on the basis of the
way the energy depends on the length. We then turn to the question 
of whether the flux tubes know about the full SU($N$) group 
or just about the $Z_N$ center and demonstrate that in fact it is the 
former. We then return to the very interesting question of whether
the excited states without an obvious Nambu-Goto interpretation
might not be revealing the excitation of new, internal degrees
of freedom. To address this question we develop a heuristic
method for comparing the shapes of the $k=2$ states with the
$k=1$ states. This study leads us to infer that all the $k=2$ energy 
eigenstates over which we do have control, are most likely to
be Nambu-Goto states with, in some cases, large corrections
to the energy dependence. Comparing those
states for which such corrections are small, with those for which
they are large, enables us to say something qualitative and striking 
about where these corrections arise in the effective string theory.
We end with a summary and our conclusions.

\section{Methodology}
\label{section_methods}

Our lattice framework is standard. We shall therefore be brief,
except with those aspects that are important for understanding 
the statistical and systematic errors of the present calculation.

\subsection{lattice field theory}
\label{subsection_lattice}

We work on periodic cubic $L_x\times L_y\times L_t$ 
lattices with lattice spacing $a$. The degrees of freedom are 
SU($N$) matrices, $U_l$, assigned to the links $l$ of the lattice.
Our partition function is 
\begin{equation}
{\cal Z}(\beta)
=
\int \prod_l dU_l 
e^{-\beta \sum_p\{ 1 - \frac{1}{N}{\mathrm{ReTr}}U_p\}}
\label{eqn_lattice} 
\end{equation} 
where $U_p$ is the ordered product of matrices around the
boundary of the elementary square (plaquette) labelled by $p$.
The continuum limit is approached by tuning $\beta \to \infty$ and 
in that limit $\beta = 2N/ag^2$ where $g^2$ is the usual continuum 
coupling, which has dimensions of mass in $D=2+1$.
Our numerical calculations use a standard heat bath/over-relaxation 
Monte Carlo, where one updates all the SU(2) subgroups of the SU($N$) 
matrices.

\subsection{calculating energies}
\label{subsection_energies}

We typically wish to calculate the energy spectrum of some sector 
of states with particular quantum numbers (spin, momentum, parity, ...).
We do so from 
correlation functions of operators $\phi(t)$ with those same 
quantum numbers. In a gauge-invariant calculation, the elementary
components of such operators are typically traces of closed loops,
where contractible loops typically project onto glueballs and
non-contractible (winding) loops typically project onto winding
loops of flux. In general we have
\begin{equation}
C(t) 
\equiv \langle \phi^\dagger(t) \phi(0) \rangle
=
\sum_{n=0}  |\langle n |\phi(0) | vac \rangle|^2 e^{-E_n t}
\stackrel{t\to\infty}{\longrightarrow}
|\langle 0 |\phi(0) | vac \rangle|^2 e^{-E_0 t}
\ \ \ \ ; \ \ \ \
E_i \leq E_{i+1}
\label{eqn_corrln} 
\end{equation}
where on a lattice we write $t = an_t$, so we see that all the
energies will be obtained in lattice units, as $a E_i$ ($n_t$ is the 
separation in the Euclidean time direction in lattice units). 
Clearly, given infinite accuracy one could calculate the whole
spectrum from one such correlation function. (Apart from states with 
accidentally zero overlap.) In the real world of finite accuracy,
however, extracting more than the ground state energy, $E_0$, is an
ill-conditioned problem so we shall proceed by the standard
variational calculation described later on.

To obtain $E_0$ from a numerical calculation of $C(t)$, 
we perform a fit with a single exponential to $C(t)$ for
$t\geq t_1$ and choose for $t_1$ the lowest value for which
we obtain an acceptable fit. This is a way to estimate
the asymptotic exponential decay in eqn(\ref{eqn_corrln}).
Now, since the fluctuations that determine the 
error in the Monte Carlo calculation of $C(t)$ are themselves
proportional to a higher order correlation function which 
can be seen to possess a disconnected piece, the
error is approximately independent of $t$, whereas $C(t)$,  as we see 
from eqn(\ref{eqn_corrln}), decreases at least exponentially with $t$.
Thus the error/signal ratio increases exponentially with $t$ and 
one can only hope to obtain $E_0$ accurately if the corresponding
ground state already dominates $C(t)$ at small values of $t$, i.e.
one needs operators $\phi$ that have a large projection onto the
desired ground state. This can be achieved using standard 
smearing/blocking techniques (see e.g. 
\cite{glued3,blmtuw-glue04}) 
which produce operators that are smooth on physical length scales,
with little computational effort. If $aE_0$ is small enough that the
error/signal ratio is small for a nontrivial range of $t=a n_t\geq t_1$,
then the success of a single exponential fit provides significant evidence
for it being appropriate. If however  $aE_0$ is so large that this range 
is small, then we lose such evidence, and typically we will be using
a single exponential fit over a range of $t$ where the contribution of
higher excited states may still be significant. Thus our estimate of
$aE_0$ will tend to become too high as $aE_0$ increases. This is
an important sytematic error that will be visible in some of our
results -- in particular when we consider very long flux tubes
(which are very massive) or highly excited flux tubes, even 
when they are not very long.

To calculate the excited states as well as the ground states,
we carry out the following variational calculation. 
Consider a set of $n$ different smeared/blocked operators,
$\{\phi_i(t) \ ; \ i=0,...,n-1\}$, with the desired quantum numbers,
and which have been normalised (but are not in general orthogonal). 
Now find the normalised linear 
combination of these operators, $\phi(t) = \Phi_0(t)$, that maximises 
$C(t=t_0) = \langle \phi^\dagger(t_0) \phi(0) \rangle
=\langle \phi^\dagger(0)| \exp\{-Ht_0\} | \phi(0) \rangle $
where $t_0$ is chosen small enough that all the correlators are accurately
determined. (In practice we choose $t_0=a$.) Then $\Phi_0$ provides 
our best estimate of the ground state wavefunctional. Now consider
that part of the basis $\{\phi_i\}$ that is orthogonal to $\Phi_0$,
and find the operator  $\phi(t) = \Phi_1(t)$  that once again maximises
$C(t=t_0) =\langle \phi^\dagger(0)| \exp\{-Ht_0\} | \phi(0) \rangle $ 
but now over this reduced basis. Then $\Phi_1$ provides 
our best estimate of the first excited wavefunctional. Repeating this
procedure we obtain a set of operators, $\{\Phi_i\ ; \ i=0,1,... \}$, that 
are our variational approximation to the exact energy wavefunctionals, 
$\{\Psi_i\}$. For each  $\Phi_i$ we now consider the correlation function 
$\langle \Phi_i^\dagger(t) \Phi_i(0) \rangle$, perform single
exponential fits as described above for the ground state, and 
extract an estimate of $aE_i$.

As for the ground state, it is critical that one has good overlaps,
and in practice that means that we need 
$|\langle \Psi_i^\dagger(0) \Phi_i(0) \rangle|^2 \geq 0.9$.
To achieve this for several excited states we need a large basis of
operators. In practice we use O(200) operators, as described below.
For excited states there are additional systematic errors. First,
the excited state may be unstable, e.g. if $(E_i - E_0)$ is
greater than the lightest scalar glueball mass. This will not lead
to any visible effects in our calculation as long as the decay width 
is small. This will be the case at large $N$, where decay widths vanish, 
but even in SU(3) they are usually too small to be important. 
(Hence the difficulty in observing string breaking directly in 
lattice QCD.) Another systematic error arises from the fact
that an excited trial operator, $\Phi_i$, may have a non-zero
overlap onto a lighter eigenstate, i.e. 
$|\langle \Psi_j^\dagger(0) \Phi_i(0) \rangle|^2 \neq 0$ for $j < i$. 
If this overlap is small then it is only at large $t$ that
$\langle \Phi_i^\dagger(t) \Phi_i(0) \rangle$ will behave like
$\propto \exp\{-E_j t\}$ and there will be an intermediate
range of $t$ where it will behave like $\propto \exp\{-E_i t\}$,
and where one can extract $E_i$. However as this overlap becomes
more significant, the ambiguity in extracting  $E_i$ obviously
will become larger. Another systematic error arises from the fact
that the excited states show near-degeneracies that can be 
distorted by mixing effects which can be enhanced and complicated by the 
effects of statistical fluctuations in our variational calculation.
\footnote{For example, a correlation between two nearly-degenerate states, 
that would average to zero with infinite statistics, can induce an 
artificial level repulsion.}
For all these reasons we shall confine ourselves 
to discussing only the lightest few excited states.

\subsection{winding flux loops}
\label{subsection_loops}

Consider a loop of fundamental flux that winds once around the 
$x$-torus, so that it is of length $l=aL_x$. A generic operator
$\phi_l$ that couples to such a periodic flux loop is the trace 
of an ordered product of link matrices, $l_p$, along a space-like 
curve $p$ that winds once around the $x$-torus, with the matrices
in the fundamental representation. (This is an example of a Polyakov 
loop, or Wilson line, which may be in a general representation.)
If we multiply all the $U_x$ link
matrices in some given $x$-slice of the 3-volume by an element 
$z\in Z_N$, then $l_p \to z l_p$. This center symmetry keeps track
of the net winding: a $k$-string will transform as 
$l_{p,k} \to z^k l_{p,k}$. We will now specialise to the $k=2$ strings
which are the subject of this paper.

We will consider $k=2$ flux tubes of two kinds. First there is
a flux tube that winds once around the torus but contains flux 
in a representation of SU($N$) that
arises from the product of 2 fundamentals and any number of adjoints.
The trace of a Wilson line in such a representation can in general be
expressed in terms of a sum of operators that involve traces of powers
and powers of traces of $l_p$ and of $l_p^\dagger$, such that 
the net power of $l_p$ minus the net power of  $l_p^\dagger$ is 2.
In practice we shall limit ourselves to the operators 
$\mathrm{Tr}l^2_p$ and $\{\mathrm{Tr}l_p\}^2$ whose linear combinations
correspond to the totally antisymmetric ($k=2A$) and totally symmetric  
($k=2S$) representations
\begin{equation}
\Phi_{2A} = \mathrm{Tr}l^2_p - \{\mathrm{Tr}l_p\}^2
\ \ \ ; \ \ \
\Phi_{2S} = \mathrm{Tr}l^2_p + \{\mathrm{Tr}l_p\}^2
\label{eqn_phi2AS} 
\end{equation} 
Since a $k=2S$ source can become a  $k=2A$ source through
gluon screening, there is no reason for such operators to
be orthogonal or to assume that such SU($N$) representations
will be useful in labelling the actual eigenstates. Whether
they are useful is a dynamical question that we shall determine in this paper.
The second type of $k=2$ flux tube that we consider is one that
carries fundamental flux and winds twice around the torus.
A typical trial operator for such a flux tube will be
\begin{equation}
\Phi_{w=2} = \mathrm{Tr}\{l_p l_{p^\prime}\}
\label{eqn_phik1w2} 
\end{equation} 
where the two single-winding paths $p$ and $p^\prime$ are in general 
not the same but the joint path $\tilde{p} = p p^\prime$ is continuous
and periodic with a period $2l_x$. Note that the $w=2$
operators need not be orthogonal to the $k=2A$ and $k=2S$ operators.
Indeed because of the strong $k=2$ binding, one might expect
that if $p$ and $p^\prime$ are similar, so that the second
winding of the flux tube is physically close to the first
winding, the attractive interactions will give a significant
probability of binding into a singly-wound $k=2$ flux tube,
and this will be reflected in the overlaps of the corresponding
operators. One has therefore the heuristic expectation that
to see a genuine $w=2$ flux tube one needs to use very different paths
$p$ and $p^\prime$.

In order to obtain good overlaps onto the true energy
eigenstates, we use smeared/blocked links in our construction
of $l_p$. Usually, and always when it matters, as in the projection 
onto specific representations of the flux, we make sure that
our blocked matrices are all SU($N$). (For historic
reasons this is not always the case for some of our oldest
calculations, although this will not affect any conclusions   
we shall draw from them.) Using operators with different levels 
of smearing probes different transverse size scales, and this 
forms part of our variational basis. 

What are the other quantum numbers that label such a flux tube? First,
we recall that in 2 space dimensions, there are no rotations around 
the axis of the tube and in the confining phase a flux tube
that is rotated by $\pi/2$, so that it winds around the $y$-torus,
will have no overlap with one winding around the $x$-torus, so
this symmetry is uninteresting. We do however have translations 
and corresponding momenta. We project onto zero-momentum transverse 
to the flux tube, $p_y=0$, by adding all $y$-translations of
our basic operator. We do not consider $p_y\neq 0$ since previous
calculations have shown that for the range of lattice spacings
we shall consider, the continuum energy-momentum dispersion
relation is accurately satisfied and so there does not seem to
be anything new that one might learn using non-zero transverse momenta.
More interesting is the momentum along the flux tube, $p_x=2\pi q/l$. 
Here we translate the loop by $x_0$ in the $x$ direction, multiply it 
by $\exp\{ip_x x_0\}$, and sum
over all such translations. Clearly if the flux tube is invariant
under translations along its axis, as we might expect if it is in
its ground state, then imposing $q\neq 0$ in this way will produce 
a null state. To have $q\neq 0$ the state must be excited in 
some way. Thus calculating the spectrum for various  $q\neq 0$
promises to teach us something new. Another symmetry is
(two dimensional) parity, $P$. Again one expects that to have $P = -$
one requires a deformation of the flux tube, i.e. an `excitation', 
that is different under reflection. Finally there is charge 
conjugation $C$. This operation changes the direction of the arrow,
$l_p \to l_p^\dagger$, so that $\mathrm{Re} \phi$ is $C=+$
and $\mathrm{Im} \phi$ is $C=-$. Since $l_p \to z l_p$ and
$l^\dagger_p \to z^\dagger l^\dagger_p$ under the center transformation
introduced earlier, there is in general zero overlap between
$l_p$ and  $l_p^\dagger$ so the  $C=+$ and $C=-$ sectors are degenerate
and the label is therefore uninteresting. However this is not the
case when $k=N/2$. In that case there can be mixing and the $C=\pm$ sectors 
need not be degenerate. In our case, that occurs for $k=2$ in SU(4).
However it is easy to see that, for $SU(4)$, $\mathrm{Im} \Phi_{k=2A}\equiv 0$.
Thus to the extent that there is a $k=2A$ sector of states, we will
only have $C=+$ therein. For $k=2S$ on the other hand,  $C=\pm$
can be a useful label in SU(4). 

In Appendix A we describe in more detail the construction of the 
actual operators we use.

\section{Flux tubes as strings}
\label{section_strings}

The immediate goal of our calculations is to provide 
information about the spectrum of $k$-strings that will
complement what we have learned previously about fundamental 
$k=1$ flux tubes. Our wider purpose, however, is to 
provide useful information for theoretical approaches to
understanding confinement and the dynamics of QCD. Thus a 
central step in our analysis is to compare our results 
to what one might expect on very general grounds,
so as to pin-point what might be new and interesting.
In this section we shall first discuss what these general
expectations might be, and then we summarise the detailed spectrum of
the free Nambu-Goto string theory (in flat space-time) since
this turns out to provide a very useful comparative benchmark.  

The flux tubes that will interest us here, are those that are wound 
around a spatial $x$-torus with all other space-time dimensions
effectively infinite. As we shrink the length of this torus, $l=aL_x$,  
such a system will undergo a phase transition at a critical value
$l=l_c=1/T_c$ where $T_c$ is the deconfining temperature. 
For $l\leq l_c$ the ground state is no longer 
a flux tube, but rather has zero energy. In $D=2+1$ and for
$N\geq 4$ one finds $l_c \surd\sigma_f \simeq 1.1$
\cite{Tcd3}.
Thus if we want to discuss closed flux tubes, we must choose $l> l_c$.

\subsection{general expectations}
\label{subsection_general}

Consider a flux tube of length $l=aL_x$ winding around the $x$-torus.
Translations leave its energy unchanged. However under transverse
translations the state is not the same: the choice of a position
for the flux tube breaks this symmetry spontaneously
and we expect to have corresponding massless fluctuations
in the excitation spectrum of the flux tube. Since we would expect the 
ground state to be invariant under translations along the flux tube, we do 
not expect this longitudinal translation symmetry to be spontaneously 
broken and we do not expect any further Goldstone bosons from that source. 
   
Now, since the SU($N$) field theory has only one scale, say
$l_\sigma = 1/\sqrt{\sigma_f}$, the width of the flux tube should in 
general be on that scale. On length scales $\gg  l_\sigma$, and
for excitation energies $\ll  1/l_\sigma$ the flux tube should
effectively behave as a thin string. The classical fluctuations of 
such a periodic string have wavelength $\lambda_n = l/n$ and hence 
frequency $\omega_n = 2\pi n/l$. The corresponding oscillators will be 
quantised, and the excitations can be thought of as made up out of phonons 
of energy $\omega_n$ and with momenta $p_n = \pm\omega_n$ along the string.
This is the expected Goldstone boson, which is massless and can
have any momentum allowed by periodicity. At large $l$ all these
eigenstates converge 
\begin{equation}
\lim_{l\to\infty} E_n/\sigma l = 1, \quad \forall n.
\label{eqn_stringy} 
\end{equation} 
At finite $l$ there will be shift in the ground-state energy from
the zero-point energies of the oscillators (essentially a Casimir energy)
and this leads to the well-known Luscher correction
\cite{LSW}
to the string energy:
\begin{equation}
E_0(l) 
\stackrel{l\to\infty}{=} 
\sigma l \left( 1 -  \frac{\pi}{6}\frac{D-2}{\sigma l^2} \right)
\label{eqn_luscher} 
\end{equation} 
which reflects the asymptotic low-lying $n=0$ excitation spectrum
\begin{equation}
E_n(l) 
\stackrel{l\to\infty}{=} 
\sigma l \left( 1 + 
\frac{4\pi}{\sigma l^2}\left(n-\frac{D-2}{24}\right) \right)
\label{eqn_exasym} 
\end{equation} 
The coefficient of the $O(1/\sigma l^2)$ term is universal and the value 
shown in eqn(\ref{eqn_luscher},\ref{eqn_exasym}) is the one for the simple 
bosonic string universality class, where the only massless excitations
of the flux tube arise from the spontaneously broken transverse
translation invariance. There is strong numerical support for 
this choice in the case of both fundamental and $k=2$ flux tubes
\cite{gs_k1,gs_k},
and this is of interest since in principle there might have been
other, less obvious, massless fluctuations that would have
led to a different coefficient. 

The large-$l$ expansion in powers of $1/\sigma l^2$, that we see
in eqns(\ref{eqn_luscher},\ref{eqn_exasym}), arises naturally in
the classic framework of Luscher, Symanzik and Weisz,
\cite{LSW,LW}
where one categorises and constrains the possible interaction terms 
in the transverse displacement fields. 
These interaction terms will in general lead to non-trivial interactions 
amongst the phonons and are encoded in the higher order corrections to
the ground and excited state energies shown in  
eqns(\ref{eqn_luscher},\ref{eqn_exasym}).
This approach has recently led to the remarkable conclusion 
\cite{LW}
that the next term in this expansion is  $\propto 1/(\sigma l^2)^2$ 
with the same coefficient as in the Nambu-Goto spectrum (see below). 
A similar conclusion was independently reached in
\cite{JD}
using the alternative Polchinski-Strominger framework 
\cite{PS}
where one builds up an effective D-dimensional string theory
within  a general string path-integral setting.

These two theoretical approaches have in common that they only
hope to treat very long flux tubes and, in effect, only eigenstates 
that are string-like in the sense of eqn(\ref{eqn_stringy}).
However it is common to think of the confining flux tube as
being qualitatively similar to a Nielsen-Olesen vortex, albeit
one that is a non-Abelian dual-superconducter. Such a vortex 
will in general also have a pattern of excitations that reflects
the effective theory producing its non-trivial structure. 
(For an explicit example in related, more tractable theories see
\cite{Tong}.)
These excitations will typically be massive, leading to
string eigenstates with energies that will typically be `much' 
larger than those of stringy excitations at large enough $l$: 
\begin{equation}
\tilde{E}(l) 
\simeq
E_0(l) + \mu
\stackrel{l\to\infty}{>}
E_n(l)
\simeq  
\sigma l + \frac{4\pi}{l}\left(n-\frac{D-2}{24}\right) + ...
\label{eqn_new} 
\end{equation} 
One's first guess for $\mu$ might be something on the order of
the lightest glueball mass, or the splitting between that state
and its first excitation. Using values for the lightest glueball masses from \cite{glued3} this would suggest
\begin{equation}
\mu \sim 2 - 4 \surd\sigma_f.
\label{eqn_mu} 
\end{equation} 
Such states should become particularly visible at smaller values 
of $l$ where the gap between the stringy states becomes larger
than this estimate for $\mu$.

One might simply give up on describing such massive excitations 
in the belief that the spectrum of anything but a very long flux
tube will be too complex to be theoretically tractable. However
our earlier calculations of the low-lying excitations of the
fundamental flux loop have revealed that the simple stringy 
predictions of the free bosonic Nambu-Goto string theory hold 
very accurately even for flux tubes as short as
$l\sqrt{\sigma_f} \simeq 1.5$
\cite{gs_k1,es_k1}.
That a flux tube which is only a little longer than it is wide 
should display simple string-like excitation modes is not
something that one can understand within a generic vortex picture.
It suggests that despite appearances short flux tubes do have a 
stringy description. Does this mean that the flux tube is indeed a 
thin string, with its apparent width simply a manifestation
of the zero-point fluctuations described above? This would
immediately come up against the difficulties of defining a consistent
bosonic string theory in 3 or 4 dimensions. A more interesting
possibility is offered by a dual gravity description of the 
AdS/CFT type. (In fact non-CFT and possibly non-AdS.)
Here the flux loop is indeed always described by
a string but now this string is in a geometry that is, at least 
initially, 10 dimensional. Of course the spectrum is now that
of a Nambu-Goto string in curved rather than flat space-time,
and this curvature is crucial to providing linear confinement rather
than a conformal-like Coulomb interaction. In this picture it may
not be unnatural for some stringy states to arise even for short
strings, and the massive modes are of great interest as they 
provide clues as to the geometry of the dual theory.

\subsection{Nambu-Goto closed string spectrum}
\label{subsection_NG}

The simplest specific bosonic string theory is the free string 
theory: Nambu-Goto in flat space-time. Of course such a theory
can only be consistently formulated in 26 dimensions. However,
as discussed by Polchinski and Strominger
\cite{PS},
if one is interested in an effective string theory for
single long strings, then the Weyl anomaly does not pose
an obstacle to working in $D < 26$. (As was already
pointed out long ago by Olesen
\cite{Olesen}.)
The spectrum of the Nambu-Goto free bosonic string theory
in $D=2+1$ or $D=3+1$ and in the sector of a single long
string, has also long been known
\cite{Arvis,Pol_book}.
We shall briefly summarise this spectrum here, since it will
provide our central point of comparison in this paper.

Consider a single closed string that wraps $w$ times around a 
spatial torus. The free string spectrum corresponds to 
quantised oscillations of the string that may be thought of as
phonon modes traveling clockwise (right-handed) and anticlockwise
(left-handed) along the background string. Our convention is to take 
the momenta of the latter as positive and of the former as 
negative. The modes are massless, corresponding to the Goldstone
bosons that arise from the fact that the presence of the background 
string spontaneously breaks the transverse translation invariance.
So the energy of a phonon equals the absolute value of its momentum.
This momentum is along the string and is constrained by the latter's 
periodicity to be $\pm 2\pi k/lw$ where $k=0, 1, 2,\dots$ and 
$lw=aL_x w$ is the length of the string. Of course, the total
momentum is constrained to be  $2\pi q/l \equiv 2\pi qw/lw$, 
with $q=0,\pm 1,\pm 2,\dots$, since the Hamiltonian is defined 
on a spatial torus of size $l$ (and not $lw$). 

Let $n_{L}(k), n_{R}(k^\prime)$ be the number of left and right
handed phonons with momentum  $2\pi k/lw$ and $- 2\pi k^\prime/lw$
respectively. A free string state is defined by the total set of
phonons and the total energy and momentum will be determined by
\begin{equation}
N_L= \sum_{k} \,\,\sum_{n_L(k)} n_L(k) \,k, \qquad
N_R= \sum_{k^\prime} \,\,\sum_{n_R(k')} n_R(k') \,k',
\label{eqn_sum_phonons}
\end{equation}
The sum of the phonon momenta is just the total momentum of
the excited string and is therefore constrained to be
\begin{equation}
N_L - N_R = qw.
\label{eqn_mom_cons}
\end{equation}
In addition the states have a parity, $P$. Under $P$ (in two
space dimensions) the transverse displacement changes sign,
so the corresponding creation operators for these displacements, the right-handed and left-handed phonon operators, will change sign.
That is to say,
\begin{equation}
P 
=
(-1)^{\mathrm{number \ of \ phonons}}
\equiv
(-1)^{\sum_{k} \,\,\sum_{n_L(k)} n_L(k)
+\sum_{k^\prime} \,\,\sum_{n_R(k')} n_R(k')}.
\label{eqn_parity}
\end{equation}
Writing down the 4-momentum of such a state and taking
its scalar product, one obtains the energy-squared:
\begin{eqnarray}
E^2_{N_L,N_R,q,w} &=& \left(\sigma \,l w\right)^2 + 8 \pi \sigma
\left( \frac{N_L+N_R}2 - \frac{D-2}{24}\right) + \left(\frac{2\pi
q}{l}\right)^2
\label{eqn_NG}
\end{eqnarray}
where the negative piece in the middle term arises from the zero-point
energies of the oscillators. These
energy levels will have a degeneracy that is determined by the number
of different ways one can produce the same value of $N_L+N_R$ in 
eqn(\ref{eqn_sum_phonons}). As an example we present in Table~[\ref{levels}]
the seven lowest energy levels of a closed bosonic string that winds 
once around the torus for $q=0,1,2$, in terms of the left and right 
movers (creation operators) $\alpha_{-k}$ and $\bar{\alpha}_{-k'}$. 

\subsection{corrections to the free string}
\label{subsection_corr}

We now consider how to incorporate corrections to the free string
spectrum. We consider $q=0$ and $w=1$ but the extension to other
values should be obvious.

The conventional Luscher analysis 
\cite{LSW,LW}
of confining strings focuses on the effective Lagrangian for the 
displacement field, and categorises the interactions in powers of 
these fields and derivatives. Originally this showed that in an
expansion of $E(l)$ in powers of $1/\sigma l^2$, the first correction,
which arises from the oscillator zero-point energies, is universal
and is proportional to the string central charge. Given the universality
it is no surprise that this term, the well-known Luscher correction,
is precisely what one obtains as a  first correction to $\sigma l$
when one expands the Nambu-Goto expression for $E(l)$ in 
eqn(\ref{eqn_NG}) in powers of $1/\sigma l^2$. More recently
\cite{LW}
this analysis has been carried further, and it has been shown, on quite
general grounds, that in $D=2+1$ the next term in this expansion also
coincides with what one obtains from Nambu-Goto, i.e.
\begin{equation}
E_{n} = \sigma l + \frac{4\pi}{l} \left(n-\frac{1}{24}\right) - 
\frac{8\pi^2}{\sigma \,l^3}
\left(n-\frac{1}{24}\right)^2 + {\cal O}(1/l^4),
\label{eqn_luscher1}
\end{equation}
where $n=0,1,2...$. Quite independently, an extension 
\cite{JD}
of the work of Polchinski and Strominger 
\cite{PS}
reached the same conclusion (but for $D=3+1$ as well). 

In our calculations of the $k=1$ flux loop
\cite{gs_k1,es_k1}
we discovered that the actual spectrum is reproduced by the free string 
Nambu-Goto spectrum much better than one would expect from 
eqn(\ref{eqn_luscher1}). In fact there is even good agreement for values 
of $l$ that are so small that the expansion of  eqn(\ref{eqn_NG}) in 
powers of $1/\sigma l^2$ completely diverges. The implication is clear 
\cite{es_k1}:
we should use the full Nambu-Goto expression as our starting point,
and consider corrections to that. In that spirit we write
\begin{equation}
E^2_n(l) = {\left( \sigma l \right)^2 + 8 \pi \sigma \left( 
n - \frac{1}{24} \right)} 
- \sigma \frac{C_p}{\left(l\sqrt{\sigma}\right)^{p}}, \qquad p\ge 3
\label{eqn_fit}
\end{equation}
where $p=3$ corresponds to the ${\cal O}(1/l^4)$ correction in
eqn(\ref{eqn_luscher1}). In this paper we will fit our $k=2$ spectra with
eqn(\ref{eqn_fit}), with the leading permitted correction of $p=3$.

\section{Spectrum of $k=2$ flux tubes}
\label{section_keq2}

We have performed calculations in SU(4) and SU(5) at $\beta=50$ and $\beta=80$ 
respectively. Since $\beta=2N^2/\lambda$, where $\lambda=g^2N$ is the 't Hooft
coupling, these values of $\beta$ should correspond to roughly the same
value of the lattice spacing $a$, and indeed we find 
\begin{equation}
a\surd\sigma_f =
\left\{ \begin{array}
{r@{\quad:\quad}l}
0.1310(2) & SU(4), \ \beta=50 \\
0.1298(1) & SU(5), \ \beta=80
\end{array} \right.
\label{eqn_sigmaf} 
\end{equation} 
Extensive experience with glueball masses
\cite{glued3}
the deconfinement temperature 
\cite{Tcd3}
and the spectrum of fundamental strings 
\cite{gs_k1,es_k1}
indicates that lattice spacing corrections for such values of $a$
are negligible and we shall therefore assume that what we obtain is
indeed continuum physics.

We calculate the $k=2$ spectra of flux tubes winding around $x$-tori
that range in length from $l=12a$ to $l=32a$. As we see from
eqn(\ref{eqn_sigmaf}) this corresponds to $l\surd\sigma_f \in
[1.56, 4.20]$. Thus our shortest flux tubes are presumably
not much longer than they are wide -- hardly `tubes' at all. For 
comparison, the minimum length $l_c$  below which we lose such flux 
tubes completely is
\cite{Tcd3}
\begin{equation}
l_c = \frac{1}{T_c} \simeq
\left\{ \begin{array}
{r@{\quad:\quad}l}
8.1a & SU(4), \ \beta=50 \\
8.3a & SU(5), \ \beta=80
\end{array} \right.
\label{eqn_lc} 
\end{equation} 
Since we are interested in accurately identifying corrections to the asymptotic
linear variation with $l$, it is important that we minimise systematic
errors that vary with $l$. Thus at small $l$ we must make $L_t$ large enough
so that the contribution of both $k=1$ and $k=2$ loops to the partition function
is negligible, and all our energies are normalised to the correct vacuum energy.
Similarly we must also make the transverse size $L_y$ large enough to avoid
finite size corrections. Thus, following a variety of checks, the lattices 
we use for SU(5) are: $12 \times 32 \times 80$, $16 \times 32 \times 40$,
$20 \times 24 \times 32$, $24 \times 24 \times 32$, $28 \times 28 \times 32$,
$32 \times 32 \times 32$. For our SU(4) calculations, which were performed 
earlier, we use slightly smaller lattices, where finite volume corrections,
for the smallest values of $l$ could be non-negligible:
$12 \times 24 \times 40$, $16 \times 24 \times 40$,
$20 \times 24 \times 32$, $24 \times 24 \times 32$, $28 \times 28 \times 32$,
$32 \times 32 \times 32$.
We typically perform 500,000 Monte Carlo sweeps and measure the full correlation
matrix every 5 sweeps.  

As described earlier, the relevant quantum numbers in labelling our $k=2$ flux
tubes are the momentum, $q$, along the flux tube and the parity $P$. 
We ignore charge conjugation since it is only relevant for $N=4$, and then
only for the heavier $k=2S$ states. In addition, the flux may be in
different representations of SU($N$). The $SU(N)$ representation is, 
however, not a conserved quantum 
number since it can change under screening by gluons. Nonetheless we shall see 
later that the mixing between $k=2A$ and $k=2S$ states is typically very 
small, and so in our analysis in this section we shall categorise the
states in this way as well. (Our basis of operators does not allow us
to explore other representations.) Note that where it significantly improves
the calculation we use our full $k=2$ basis of operators, although we
categorise the states by their dominant $k=2A$ or $k=2S$ component.

\subsection{$k=2A$}
\label{subsection_keq2A}

In Figs.~\ref{SU42Ars} and~\ref{SU52Ars} we plot the energies of the 
lightest states with 
momenta $q=0,1,2$ along the flux tube, for SU(4) and SU(5) respectively. 
We observe that for all values of $l$ the lightest $q=0$ state has $P=+$,
the lightest $q=1$ state has $P=-$, and for $q=2$ there are two nearly 
degenerate lightest states, one with $P=+$ and one with $P=-$. 
We note that these quantum numbers and (near-)degeneracies are precisely 
those of the Nambu-Goto free string theory. (See Table~\ref{levels}.)  
All these states have a projection onto $2A$ that is very much larger than 
onto $2S$ and we therefore regard them as constituting part of the $2A$ spectrum.
We have removed all lattice units by normalising the loop energies and the
loop lengths to the value of $a\surd\sigma_f$ that has been obtained 
from $k=1$ flux loops in the same calculation. (The errors on this quantity 
are much smaller than all our other errors.) 

We fit the $q=0$ ground state using the Nambu-Goto
expression with the leading  permitted correction, as in eqn(\ref{eqn_fit}),
\cite{LW,JD}, 
\begin{equation}
E^2(l) = (\sigma_{2A} l)^2 
\left(1 - \frac{\pi}{3}\frac{1}{\sigma_{2A} l^2} \right)
- \sigma_{2A} \frac{C_3}{\{\sqrt{\sigma_{2A}} l\}^3}
\label{eqn_loop2A} 
\end{equation} 
to obtain 
\begin{eqnarray}
a^2\sigma_{2A} = 0.023233(51) = 1.354(5)a^2\sigma_f \ ; \   C_3 = 2.93(21)   \qquad : SU(4) \\
a^2\sigma_{2A} = 0.025825(50) = 1.533(4)a^2\sigma_f \ ; \   C_3 = 2.28(22)   \qquad : SU(5)
\label{eqn_sig2AN4N5} 
\end{eqnarray} 
Note that the values of $\sigma_{2A}/\sigma_f$ are close to 
the quadratic Casimir ratios, $k(N-k)/(N-1)$.

We observe that when expressed in terms of natural units, as in 
eqn(\ref{eqn_loop2A}), the correction to Nambu-Goto has a
coefficient $C_3 \sim O(1)$. This is a natural value if we regard 
the string theory as an effective theory, in the Nambu-Goto 
universality class. It is to be contrasted with the value
$C_3  \sim O(1/10)$ that one finds for the fundamental flux tube,
and serves to emphasise how remarkably well the latter is
described by the free bosonic Nambu-Goto string theory. 

Using these calculated values of $\sigma_{2A}$,
we calculate the lightest $q=0,1,2$ energies of the Nambu-Goto model,
and plot the results in Figs~\ref{SU42Ars} and ~\ref{SU52Ars}. We 
observe an excellent agreement between our numerical results and the 
Nambu-Goto values. Note that once the $q=0$ ground state has
been fitted, the  $q=1,2$ energies are parameter-free predictions.

Since the Nambu-Goto prediction appears to work so well, it is
worth examining the results more closely to try and isolate how
well the phonon excitation energy is being reproduced by our
numerical results. To do this we note from eqn(\ref{eqn_NG}) 
that the piece of the energy that is due to the phonons is given by
\begin{equation}
\Delta(q,l)
=
E_{gs}^2(q;l) - E_{gs}^2(q=0;l) 
- \left ( \frac{\pi q}{l}\right )^2
\label{eqn_exq} 
\end{equation} 
Using  our numerically calculated values for $E_{gs}(q;l)$
we obtain the results displayed in Figs.~\ref{SU4sub2A}
and ~\ref{SU5sub2A} for SU(4) and SU(5) respectively.
(We choose to use our numerical value for $E_{gs}(q=0;l)$ 
rather than the Nambu-Goto prediction since the former already
has some small corrections which we would regard as belonging
to the Casimir energy rather than to the excitation energy 
that we are trying to isolate here.) For comparison
we show the Nambu-Goto phonon excitation energies as well.
We observe that the simple free string excitation energies 
provide an excellent description of the string excitation
energies over almost our whole range of string lengths, with
significant deviations only beginning to appear at the
smallest value of $l$, $l\surd\sigma \simeq 1.5$. 
Given the qualitative (quantum numbers, degeneracies) and
quantitative agreement between Nambu-Goto and our results,
we will assume that these eigenstates are very close to those
of the free string theory, with the corresponding phonon
excitations.

We began this Section with the ground states for various $q$ because, 
as explained earlier, ground state energies can be extracted more 
reliably than excited state energies. We now turn to the excited
states that form the low-lying $q=0$, $P=+$ spectrum. We plot our 
results for SU(5) in Fig.~\ref{spectrum2AP+SU5} together with the 
Nambu-Goto predictions. (The SU(4) results are essentially the same 
and are given in Fig.~\ref{spectrum2AP+SU4}.)
The first excited state shows substantial deviations from
the free string energies, and the next two states even more so.
However at least the former does seem to be approaching
the free string energy as $l$ increases. This might tempt us to
see it as a string excitation, but this would be dangerous,
since we also observe that to
quite a good approximation this first excited state is
greater than the ground state by a constant value,
\begin{equation}
E_1(q=0,l) \simeq E_0(q=0,l) + \epsilon  
\qquad ; \quad  \epsilon \simeq 2\surd \sigma_f,
\label{eqn_delE} 
\end{equation} 
which is reminiscent of our expectation for non-stringy states as described 
in eqns(\ref{eqn_new},\ref{eqn_mu}). Such a state would cross the
level of the first string excitation at some larger $l$, and 
perhaps that, not a convergence, is what we are seeing in 
Fig.~\ref{spectrum2AP+SU5}.
At this stage we clearly cannot decide how to classify this excited 
state. We shall return to this question in a later Section, where
we shall find that looking directly at the wavefunctional
will give us an unambiguous answer.

\subsection{$k=2S$}
\label{subsection_keq2S}

As we shall see later on, all the states we analyse appear to transform
primarily as $k=2A$ or as $k=2S$ (or as $w=2$ fundamental). However 
the lowest $k=2A$ states are much lighter 
than the corresponding $k=2S$ states, so the latter
will be much more susceptible to the systematic errors discussed in
Section~\ref{subsection_energies} -- in particular to their masses 
being over-estimated. 

Bearing this caveat in mind, we display in Figs.~\ref{SU42Srs} 
and ~\ref{SU52Srs} the energies 
of the lightest  $k=2S$ states with momenta $q=0,1,2$, for SU(4) 
and SU(5) respectively. Qualitatively the picture is much as for the
$k=2A$ states: we find that for all values of $l$ the lightest 
$q=0$ state has $P=+$, the lightest $q=1$ state has $P=-$, 
and for $q=2$ there are two nearly degenerate lightest states, one with 
$P=+$ and one with $P=-$. All this fits in with the Nambu-Goto free
string picture. Turning to a quantitative comparison with Nambu-Goto,
we encounter an immediate difficulty. Our estimate of the $l=32$ ground 
state energy is too high to fit in with an $E(l)$ tending to an
asymptotically linear behaviour. We therefore assume that what we
are seeing here is the result of a systematic error of the kind described 
in Section~\ref{subsection_energies}, and which is perhaps twice the 
size of the statistical error.
If we exclude this value from our fit to the $q=0$ ground state using
eqn(\ref{eqn_NG}), we obtain  
\begin{eqnarray}
a^2\sigma_{2S} = 0.03957(45) = 2.306(27)a^2\sigma_f    \qquad : SU(4)\\
a^2\sigma_{2S} = 0.03793(33) = 2.251(20)a^2\sigma_f    \qquad : SU(5)
\label{eqn_sig2SN4N5} 
\end{eqnarray} 
(where the correction coefficients, $C_3$, are not well-determined).
In fact if we had included the $l=32$ data point we would have obtained almost
identical values of $\sigma_{2S}$, but with a much worse $\chi^2$. 
We note once again that the values of $\sigma_{2S}/\sigma_f$ are close to 
the quadratic Casimir ratios, in this case $k(N+k)/(N+1)$.

Using these values of  $\sigma_{2S}$, we show the corresponding $q=0,1,2$ 
ground state Nambu-Goto spectra in  Figs.~\ref{SU42Srs} and ~\ref{SU52Srs}. 
Semi-quantitatively the agreement is good; however the longer, more massive
states tend to overshoot the predicted values. It is plausible that this
is again due to the systematic errors of our calculation growing with
the calculated mass.

In  Fig.~\ref{SU5sub2S} we extract the phonon excitation energies from 
our SU(5) values, using eqn(\ref{eqn_delE}). (The SU(4) results are 
very similar.) There is good qualitative agreement with 
Nambu-Goto, but clearly we cannot say more than that.

In contrast to the $k=2A$ case, we have severe difficulty in extracting 
excited state energies and hence do not attempt to produce a plot that is
analogous to Fig.~\ref{spectrum2AP+SU5}.

To conclude, it is probably fair to say that the $q=0,1,2$ ground states
provide significant evidence that the $k=2S$ flux tube also behaves
like a free bosonic string, at least for these states, but that any attempt
at a stronger statement is hampered by the current uncertainties of our
numerical calculations for these very massive excited states.
  
%
%
%
%
\section{Group or centre?}
\label{section_sunorzn}

For each $k$, and for each set of conserved quantum numbers $P,C$ and $q$,
there should only be one absolutely stable flux tube. (Except e.g. where
it can decay into a string  with the same $k$ but with some of the other
quantum numbers different, and accompanied by glueballs that carry the compensating 
quantum numbers.) However, flux tubes carrying flux in an $SU(N)$ 
 representation with $N$-ality $k$ that is, in general, not stable, can be 
stable enough to be readily identifiable in our calculations. If so it is an interesting
question to ask if they are individually described by an effective
bosonic string theory with an appropriate string tension. 
We have, of course, already answered this question in the affirmative
in earlier sections, where we examined the $k=2$ flux tubes in the
totally symmetric and antisymmetric representations. (Which are
the only ones accessible to our limited choice of operators.)
Here we will present evidence that the portion of the $k=2$ 
spectrum accessible to us is, to a good approximation, the sum of 
disjoint $k=2A$ and $k=2S$ spectra. This provides the promised 
continuation of the discussion of this question in
\cite{gs_k}
to which we refer the reader for further background.

\subsection{operator overlaps}
\label{subsection_opoverlap}

We begin by looking at the overlaps of our $k=2A$ basis operators,
$\{\Phi^i_{2A}, i=0,1,2,...\}$, onto their  $k=2S$ counterparts,
$\{\Phi^i_{2S}, i=0,1,2,...\}$. For this purpose we define 
the normalised overlaps:
\begin{equation}
O_{AS}(i,j;t)=\frac{<\Phi^{\dagger}_{2A,i}(t) 
\Phi_{2S,j}(0)>}{<\Phi^{\dagger}_{2A,i}(t) 
\Phi_{2A,i}(0)>^{1/2}<\Phi^{\dagger}_{2S,j}(t) 
\Phi_{2S,j}(0)>^{1/2}}
\label{eqn_OAS} 
\end{equation}

In Table 4 of
\cite{gs_k}
we presented results for such overlaps, at $t=0$, using a small basis of 
blocked/smeared Polyakov loops of length $l=24a$ on a $24^2 32$ lattice 
in SU(5) at $\beta=80$ (corresponding to $l\sqrt{\sigma_f} \sim 3$). 
We found that most of the overlaps were zero within errors, except 
for those where at least one of the operators was at the largest
blocking level; but even there the largest overlap was still a very
small $\sim 0.03$. This provided some striking evidence
that the theory does indeed break up into almost disjoint
$k=2A$ and $k=2S$ sectors. However, while this small basis 
has a reasonably good overlap onto the $k=2A$ and $k=2S$ ground 
states, the overlap onto excited states is in general poor.
It is therefore interesting to see what happens when we 
consider the much larger operator basis of the present paper,
which has a good overlap onto most of the low-lying states.
Another question is how much the results of
\cite{gs_k}
are affected by finite volume corrections. The point being
that the only significant overlaps were for the largest 5'th
blocking level, where the blocked link is of length $16a$ 
and has a width that is at least as large -- that is to say,
the operator more-or-less covers the whole spatial torus.
Finally a very interesting question is what happens for 
$t\neq 0$, since this probes more directly the represntation
content of the energy eigenstates.

To address the question of finite volume corrections
we perform some $t=0$ calculations in SU(4) at $\beta = 32$.
(Since in SU(4) the $k=2A$ operators are necessarily
$C=+$, we take their overlaps with the $C=+$ piece
of the $k=2S$ operators.)
We start with $l=16a$ because this also corresponds to
$l\sqrt{\sigma_f} \sim 3$. Note that on a $16^2$ spatial lattice
the 5'th blocking level really does cover the whole spatial torus;
so any finite size corrections should be larger than in
\cite{gs_k}.
We show a selection of the overlaps on a  $16^2 20$ lattice
in Table~\ref{table_OAS}, including all those which are large enough
to be significant. (Nearly all those not shown are zero within 
errors.) We see that a few overlaps are indeed not small,
but that they again occur when one of the operators is at the
largest blocking level. 
We now see what happens if we keep the same 
loop length, but vary the transverse spatial size, i.e. we
work on $16\times l_y$ spatial lattices where we vary $l_y$.
What one finds is that the overlaps rapidly decrease with
increasing $l_y$ (and rapidly increases as we decrease $l_y$
below $l$).We show examples in Table~\ref{table_OAS} for
$l_y=20$ and $l_y=48$. The decrease is however not to
zero but to a finite value; in fact the $l_y=48$ values
are asymptotic within errors. These values are very small,
and much smaller than on the original $16^2$ torus.
We now turn to what happens as we increase $l$ while staying
on an $l^2$ lattice. As we see in  Table~\ref{table_OAS}
this also strongly decreases the overlaps. In this case
all the overlaps appear to tend to zero as $l$ increases.
Of course higher blocking levels will become possible at 
larger $l$, and we would expect the largest of these to 
sometimes generate significant overlaps, but such operators 
will, eventually, have very small overlaps onto the low-lying
flux loop states. 

\begin{figure}[!htb]
\vskip -2.5cm
\begin{center}
\scalebox{1.3}{\input{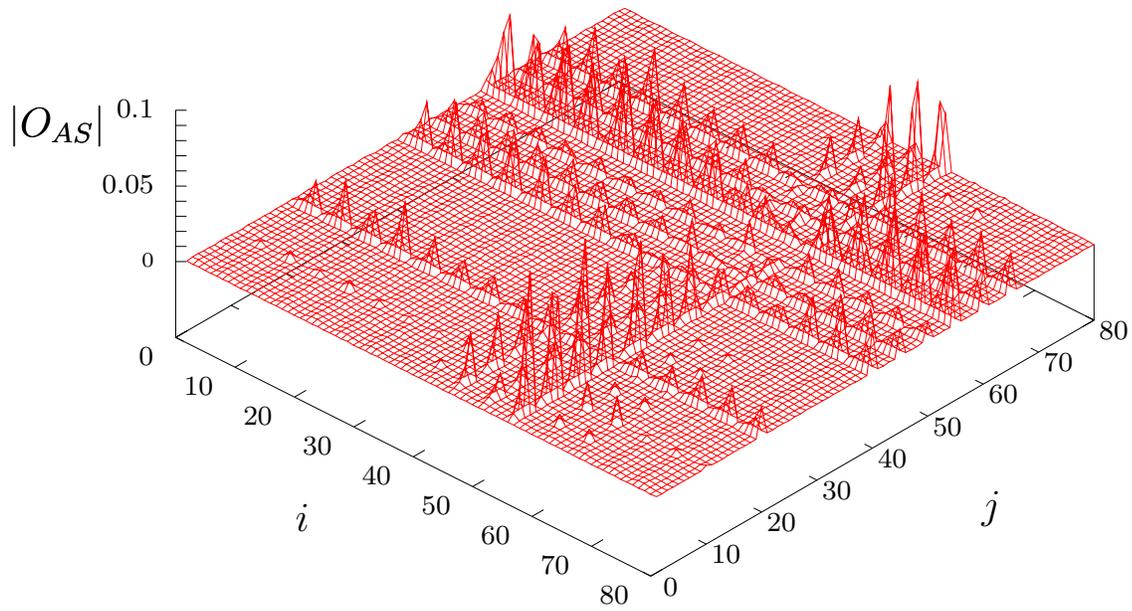} }
\end{center}
\vskip -1cm
\caption{The overlap $|O_{AS}(i,j;t=0)|$ as defined by eqn(\ref{eqn_OAS})
for $SU(4)$, $L=32a$, $P=+$ and $q=0$ with $i,j=1-80$.}
\label{Oas}
\end{figure}

We now turn to the overlap between all the operators in our extended 
basis. These  have been calculated on a $32^3$ lattice for $P=+$, 
$q=0$ and SU(4) at $\beta=50$. A Table is not practical, so we
provide in Fig.~\ref{Oas} a three-dimensional plot of the values 
of $|O_{AS}(i,j;t=0)|$ as $i$ and $j$ run over our basis of some 80 
operators. As we might have expected from our more limited study above,
the overlaps are mostly very small, and usually zero within the errors.
Only those that involve operators at the very largest blocking levels
are significant although even these are numerically small. 
(Here the 5'th blocking level is the largest, and it appears in our 
operator labeling at multiples of 5.) We can further assume, 
given what we found above, that most of these larger values are 
much enhanced by finite volume corrections.
Thus this study confirms that even when we extend the basis so that
it is quasi-complete for the low-lying spectrum of the $k=2$ flux tube,
the near-orthogonality of the $k=2A$ and  $k=2S$ sectors is
maintained. Physically this suggests that the effects of screening 
$k=2S$ down to $k=2A$ are very weak, and why that should be so, even in
SU(4), needs to be better understood.

Of course the near orthogonality of the $k=2A$ and  $k=2S$ sectors
does not necessarily imply that the energy eigenstates belong
to one sector or the other (although it is a condition for that
to be possible). To be sensitive to the eigenstate content we need
to look at $|O_{AS}(i,j;t)|$ for $t\neq 0$. Since the correlators
decrease exponentially with $t$, we only have accurate results
for the lowest values of $t$ and we will therefore restrict ourselves
in the following dicussion to $t \leq 2$. Now, if the eigenstates
were linear combinations of comparable amounts of $k=2A$ and  
$k=2S$, then we would expect to find $|O_{AS}(i,j;t\neq 0)|\sim O(1)$.
If on the other hand $|O_{AS}(i,j;t\neq 0)|$ is `small' then this suggests
that the lightest states are largely in one sector or the other.
To determine which is the case, we turn to our high statistics calculation
on a $24^2 32$ lattice in SU(5) at $\beta=80$. In Table~\ref{table_OASt}
we show the same selection of overlaps as in Table~\ref{table_OAS},
but now for $t=0,1,2$. (The $t=0$ results differ from those shown in
\cite{gs_k}
because the statistics here are ten times greater.) We see that the $t\neq 0$
overlaps do indeed remain very small. Moreover some lower statistics 
calculations on a $24\times 32 \times 32$ lattice show that with such a
larger transverse size, the overlaps decrease by a factor $\sim 5 - 10$.
Thus we conclude that the overlaps for $t \neq 0$ are also very small,
and the low-lying eigenstates cannot be comparable mixtures of
$k=2A$ and  $k=2S$. To make this more precise we shall now turn to an 
analysis of the actual eigenstates.

\subsection{state overlaps}
\label{subsection_stateoverlap}

What we will do in this Section is to estimate the overlaps
of the actual string eigenstates onto the $k=2A$ and $k=2S$ sectors.
Our approach to this question, which is somewhat different to that 
employed in
\cite{gs_k},
is as follows. 

Let $\{\tilde{\Psi}_i\}$ be the true energy 
eigenfunctionals ordered, as usual, by their energy $\{E_i\}$.
Using only our $k=2A$ basis, let the variationally calculated trial
eigenfunctionals be $\{\tilde{\Phi}_{2A,i}\}$. Suppose now that
$\tilde{\Phi}_{2A,j}$ corresponds to $\tilde{\Psi}_i$, where 
$j \leq i$. (The possibility $j<i$ arises when there is an eigenstate 
$\tilde{\Psi}$ with a small enough overlap onto our $k=2A$ basis, 
that it does not lead to a corresponding $\tilde{\Phi}_{2A}$,
so that the set of variational trial eigenfunctions is incomplete.)
Then the normalised correlation function of 
$\tilde{\Phi}_{2A,j}$ will be $\simeq \gamma^2_{2A} \exp\{-E_i t\}$
in some range of $t$ -- the effective energy plateau. Then
$\gamma_{2A}$ gives us the overlap of the eigenstate $\tilde{\Psi}_i$  
on $\tilde{\Phi}_{2A,j}$ and hence, to a good approximation, on
our $k=2A$ basis. Doing the same with the  $k=2S$ basis should
give us the corresponding overlap $\gamma_{2S}$. Now returning to 
reality, we know that we will not see such a state at all in any
sector where its overlap is small. So in practice we can usually
only calculate either $\gamma_{2A}$ or $\gamma_{2S}$ this way, but
not both. So suppose it is $\gamma_{2A}$ that is large enough to be
clearly calculable. Then instead of looking within the $k=2S$ basis 
to estimate  $\gamma_{2S}$, we look at the joint basis
$\{k=2A\}\bigoplus\{k=2S\}$. Here the overlap is $\gamma_{2AS}$
and must be larger than $\gamma_{2A}$, so the state will certainly
be visible to the variational calculation. If we take the
$k=2A$ and $k=2S$ bases to be approximately orthogonal to each other,
which we have shown above to be the case,
then we can estimate $\gamma^2_{2S}\simeq \gamma^2_{2AS}-\gamma^2_{2A}$.
If the state is more $k=2S$ than $k=2A$ then we follow the same 
procedure but with $2A$ with $2S$ interchanged.
So in this way we obtain an estimate of $\gamma^2_{2A}$ and
$\gamma^2_{2S}$ for each energy eigenstate. These estimates
are of course approximate. In particular, our $k=2A$ and $k=2S$ 
bases are not complete, so that $\gamma^2_{2A}+\gamma^2_{2A} < 1$
and becomes more so (typically) for higher excited states.
In addition the non-orthogonality of the $k=2S$ and $k=2A$ sectors
introduces an ambiguity in interpreting the results, but this is
very small.

\begin{figure}[!htb]
\centerline{\scalebox{0.65}{\input{newplotP+Q0.tex} \ 
\put(-290,250){\large $l \sqrt{\sigma_f}\sim2.1$}
\put(-105,250){\large $l \sqrt{\sigma_f}\sim3.1$}
\put(-133,86){$\parbox{0.5cm}{\rotatebox{0}
{\includegraphics[width=0.5cm]{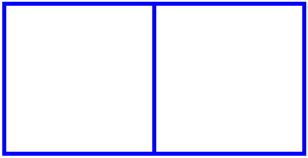}}}$}
\put(-30,86){$\parbox{0.5cm}{\rotatebox{270}
{\includegraphics[width=0.5cm]{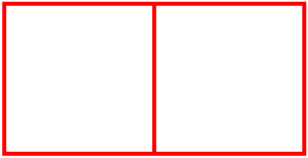}}}$}
\put(-313,56){$\parbox{0.5cm}{\rotatebox{0}
{\includegraphics[width=0.5cm]{symmetrictab.eps}}}$}
\put(-212,56){$\parbox{0.5cm}{\rotatebox{270}
{\includegraphics[width=0.5cm]{antisymmetrictab.eps}}}$}
\hspace{10mm} }
\scalebox{0.65}{\input{newplotP-.tex}
\put(-290,250){\large $l \sqrt{\sigma_f}\sim3.1$}
\put(-105,250){\large $l \sqrt{\sigma_f}\sim3.1$}
\put(-133,125){$\parbox{0.5cm}{\rotatebox{0}
{\includegraphics[width=0.5cm]{symmetrictab.eps}}}$}
\put(-31,125){$\parbox{0.5cm}{\rotatebox{270}
{\includegraphics[width=0.5cm]{antisymmetrictab.eps}}}$}
\put(-313,166){$\parbox{0.5cm}{\rotatebox{0}
{\includegraphics[width=0.5cm]{symmetrictab.eps}}}$}
\put(-220,166){$\parbox{0.5cm}{\rotatebox{270}
{\includegraphics[width=0.5cm]{antisymmetrictab.eps}}}$}}}
\vskip -0.3cm
\caption{Overlaps squared onto $k=2A$ (red circle) and $k=2S$ (blue square)
of the low-lying $k=2$ flux tube states with the lengths $l$ shown. 
In sectors, from left to right, $\{P=+,q=0\}$, $\{P=+,q=0\}$, 
$\{P=-,q=0\}$, $\{P=-,q=1\}$. All in SU(4) at $\beta=50$.} 
\label{Towers}
\vskip -0.3cm
\end{figure}

We show a selection of such analyses in Fig.~\ref{Towers}.
The examples are all from SU(4) at $\beta=50$. (The SU(5) results
would be very similar.) We show results for the $P=+, q=0$
spectrum for the two string lengths shown, as well as results
for the longer string length for the  $P=-, q=0$ and $P=-, q=1$
sectors. For each state we show our estimate of the $k=2A$ 
(red circle) and  $k=2S$ (blue square) overlap squared. We see 
that all the states are predominantly in one sector or the other.
In some cases the sum of the overlaps is significantly less than
unity, and that simply reflects the overlap of that state on our 
whole basis. This effect becomes stronger for the higher 
excited states.

\section{{\bf{$\omega=2$}} `unbound' fundamental flux loops} 
\label{section_weq2}

So far we have focused on $k=2$ bound states. However the $k=2$
sector should also include states that consist of a fundamental
flux loop that winds twice around the spatial torus, and whose
spectrum will be given, to some first approximation, by 
eqn(\ref{eqn_NG}) with $\omega = 2$. These states can be thought of as corresponding to two unbound fundamental flux-tubes. Such states might suffer large
corrections because in $D=2+1$ such a double winding loop 
will necessarily cross itself and we know that there are 
interactions between loops that are strong enough to bind them
into the $k=2A$ strings. Moreover the $\omega = 2$ loops should 
typically be more massive than the corresponding $k=2A$ states 
and so they will typically appear in a part of the $k=2$ spectrum where
they might be confused with other excited states.
All this should make them non-trivial to observe and so
what we shall attempt in this Section is no more than a preliminary
search for such $\omega = 2$ flux loops.

For this purpose we include in our basis additional operators of 
the form in eqn(\ref{eqn_phik1w2}). We then perform the same calculation 
as in the Section~\ref{section_keq2} but within this extended basis.
Comparing the resulting spectra to those previously obtained in
Section~\ref{section_keq2}, we attempt to identify new states. 
The new basis of operators that has been used in our calculation 
consists of approximately $\sim 240$ operators.

\begin{figure}[!htb]
\centerline{
\scalebox{0.9}{\input{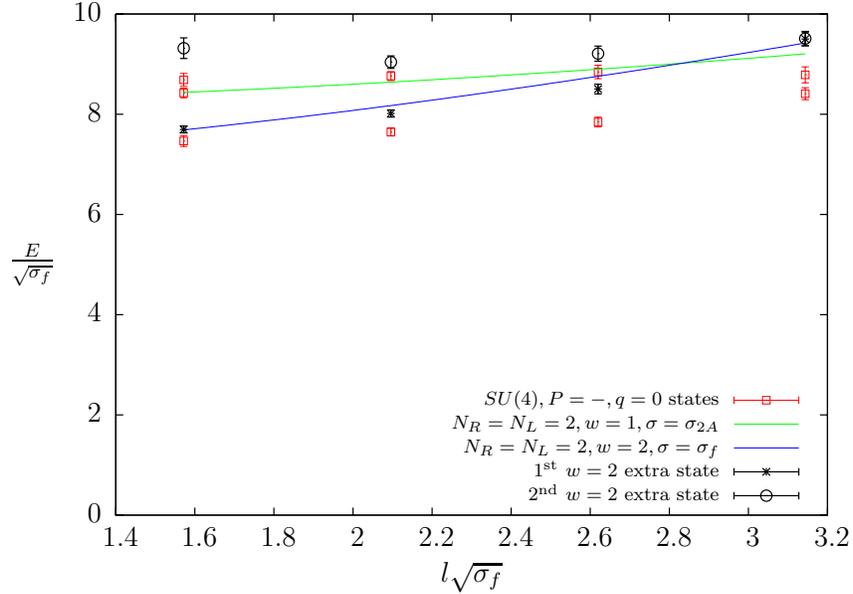} }}
\caption{Spectrum of $k=2$ flux loops with $P=-$ and $q=0$, as extracted 
using the extended basis of operators, that includes $w=2$ $k=1$ operators 
as in eqn(\ref{eqn_phik1w2}).}
\label{Extra2}
\end{figure}

\begin{figure}[!htb]
\centerline{
\scalebox{0.9}{\input{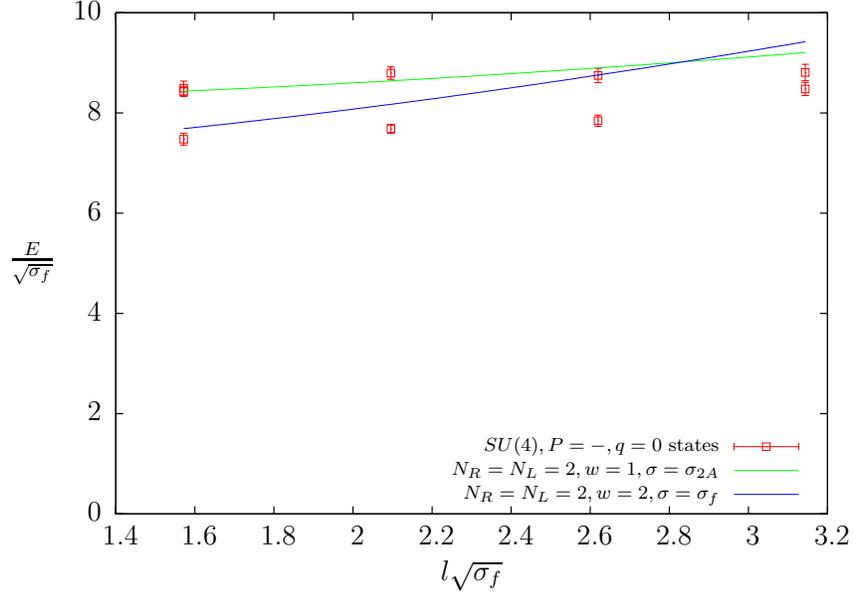} }}
\caption{Spectrum of $k=2$ flux loops with $P=-$ and $q=0$, as extracted 
using only our usual normal $k=2$ basis of $\{k=2A\}\bigoplus\{k=2S\}$ operators,
as in eqn(\ref{eqn_phi2AS}), and excluding the $\omega=2$ states}
\label{Extra}
\end{figure}

In Figs.\ref{Extra2}~and~\ref{Extra} we compare the $SU(4)$, $P=-$, 
and $q=0$ spectra obtained with and without the extra $\omega =2$
operators, and we do indeed observe that the former appears to
contain two extra states, which we have shown as solid black
symbols. For comparison, the blue line represents the NG prediction 
for a $\omega=2$ $k=1$ string, using $\sigma=\sigma_f$ and $N_R=N_L=2$ 
in eqn(\ref{eqn_NG}). The red line represents the NG prediction for the 
$\omega=1$ $k=2$ antisymmetric representation with the 
same values of $N_L$ and $N_R$ and with $\sigma=\sigma_{2A}$. 
As we see from Table~\ref{levels}, the ground state of a winding string 
with $P=-$ and $q=0$ should be two-fold degenerate, and previous 
calculations for $\omega=1$  $k=1$ flux tubes 
\cite{es_k1}
have shown that what one gets is one state which is close to the theoretical 
NG prediction and another one that approaches the theoretical 
prediction from above as $l$ increases. We observe precisely such 
a behavior in Figure~\ref{Extra2}, with 
the lighter state, represented by a star, being close to the $w=2$ 
prediction and the heavier state, represented by a circle, 
approaching the blue line from above.

\begin{figure}[!htb]
\centerline{
\scalebox{1.0}{\input{ploteig2.tex} }
\put(-136,20){{\line(0,1){9}}}
\put(-137,20){{\line(0,1){9}}}
\put(-137,20){\vector(-1,0){182}}
\put(-136,20){\vector(1,0){125}}
\put(-100,22){\scriptsize $w=2$ Operators}
\put(-280,22){\scriptsize Ordinary $k=2$ Operators}
\put(-182,194){\vector(3,-1){65}}
\put(-306,190){\includegraphics[height=1.5cm]{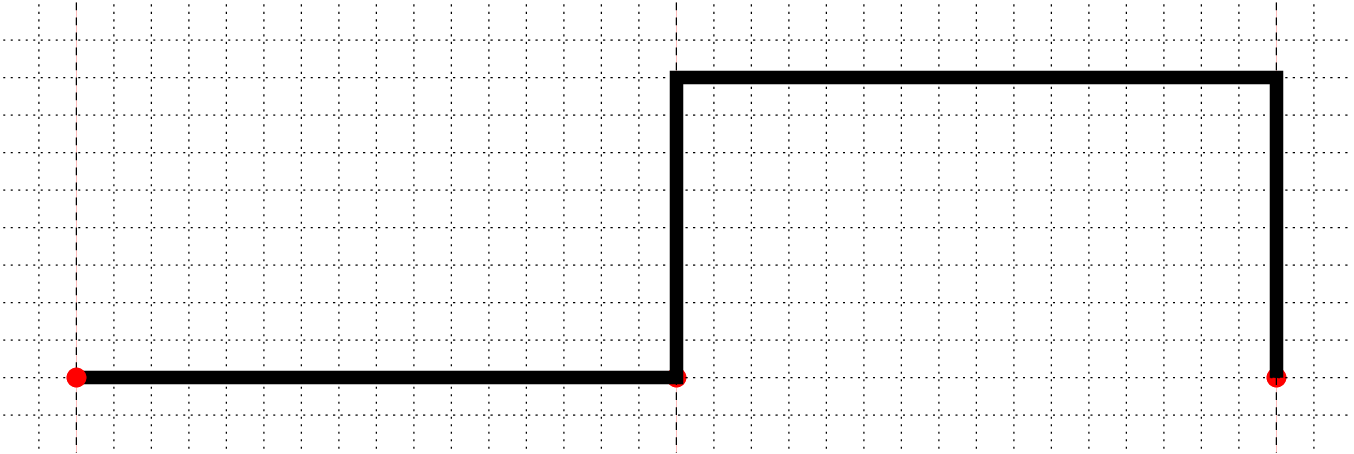}}
\put(-298,184) {\line(0,1){3}}
\put(-243,184) {\line(0,1){3}}
\put(-298,184) {\line(1,0){55}}
\put(-241,184) {\line(0,1){3}}
\put(-186,184) {\line(0,1){3}}
\put(-241,184) {\line(1,0){55}}
\put(-280,177) {\scriptsize $L=16a$}
\put(-228,177) {\scriptsize $L=16a$}}
\caption{The contribution of each of our extended basis of operators 
to the ground and first excited states shown in Fig.~\ref{Extra2}
for $L=16a$. For $i\le 100$ the operators are the ordinary $k=2$ ones and 
for $i>100$ the operators are the additional $w=2$ ones.}
\label{eigen}
\end{figure}

To confirm that these two states are primarily associated with
the extra $\omega =2$ operators in our extended basis,
we show in Figure~\ref{eigen} the projection in of the variational 
eigenstates corrsponding to the  ground and first excited states 
onto all the operators in our basis.
We have chosen to look at $l=16a$ ($l \sqrt{\sigma_f}\sim2.1$)
where these states are (supposedly) the lightest $\omega=1$ $k=2A$
and $\omega=2$  $k=1$ states. For this lattice size we have 160 
operators in total, of which the first 100 are our ordinary $k=2$
operators, and the last 60 are our extra $\omega =2$ operators.
As can be seen, the ground state projects mainly onto ordinary $k=2$ 
operators, while the first excited state projects almost entirely
onto the $w=2$ sector of the operator basis. This supports
our earlier interpretation of these states.
 
It is interesting to note that the $w=2$ operator that contributes the 
most to the first excited state is the rectangular pulse 
(Table~\ref{Operators}, number 3) at blocking level 4 and, therefore, 
the transverse deformation has a physical length of 16 lattice spacings. 
A pictorial representation of this operator is included on the panel of 
Figure~\ref{eigen}. We see that it is shaped like an oscillating closed 
string with two nodes, doubly wound around the torus. This provides
further heuristic support for our conclusion that
the complete $k=2$ string spectrum includes unbound 
wave-like $\omega =2$ states.

\section{Looking for extra massive states}
\label{section_extra}

As we remarked in Section~\ref{subsection_keq2A}, the interpretation of 
the first excited state in the $k=2A$, $P=+$, $q=0$ spectrum 
shown in Fig.~\ref{spectrum2AP+SU5}, is ambiguous. It might be the first
string excitation, with a large correction to the Nambu-Goto free string 
energy, which is shown in Fig.~\ref{spectrum2AP+SU5} and which it 
appears to be approaching as $l\uparrow$. Alternatively it might
be a new excitation that arises from one of the massive modes of
the ground state flux tube being excited. Indeed, as we see from 
Fig.~\ref{spectrum2AP+SU5}, the energy of this state does differ
from that of the ground state by roughly a constant as $l$ varies. 
In this scenario, the energy approaches the energy of the
first excited Nambu-Goto state, only to cross it at some
larger value of $l$ and in that case it is presumably the next state 
up in energy which is the true stringy excitation (and which indeed
is much closer to the Nambu-Goto value for most of our values of $l$).

Since the identification of such extra states would (begin to) provide 
a completely new source of useful information about the string
description of the gauge theory, resolving this ambiguity is
important, and that is what we turn to now.

If the state is indeed an approximate Nambu-Goto-like string excitation
of the $k=2A$ flux tube, then we would expect its wave-functional
to have the appropriate `shape'. What that `shape' should be, in
terms of our highly blocked/smeared link matrices, is not at all 
evident, but it is something we do not need to know. 
It is enough to note that we have already established 
\cite{es_k1}
that the relevant part of the fundamental $k=1$ flux tube spectrum is 
Nambu-Goto-like. Indeed the $k=1$ analogue of  Fig.~\ref{spectrum2AP+SU5}
shows only small corrections to the free string energies. So all
we need to do is to compare the wavefunctionals of the first 
excited $k=1$ and $k=2A$ states, in exactly the same lattice
calculation, and see if they are very similar or very different. 

The way we make this comparison is as follows. Let 
$\{\phi_i; i=1,...,n_o\}$ be our set of winding operators 
where the number is typically $n_o \sim 100$. In the present case, 
we choose them to have $P=+$ and $q=0$. These operators are group 
elements (not yet traced) and may be in any representation of SU($N$).
Suppose the flux is in the representation ${\cal R}$. When we
perform our variational calculation over this basis, we obtain
a set of wavefunctionals, $\Phi^n_{\cal R}$, which are an
approximation to the corresponding eigenfunctions of the
Hamiltonian. These wavefunctionals are linear combinations of 
our basis operators:
\begin{equation}
\Phi^{n}_{\cal R}
=
\sum^{n_o}_{i} b^{n}_{{\cal R},i} c_{{\cal R},i}{\rm Tr}_{\cal R} (\phi_i)
\equiv
\sum^{n_o}_{i} b^{n}_{{\cal R},i} {\rm Tr}^{\prime}_{\cal R} (\phi_i)
\label{phie2}
\end{equation}
where the coefficients $c_{{\cal R},i}$ are chosen so that
we have the normalisation condition
\begin{equation}
\langle {\rm Tr}^{\prime\dagger}_{\cal R} (\phi_i(0))
{\rm Tr}^{\prime}_{\cal R} (\phi_i(0))\rangle
= 1
\label{phinorm}
\end{equation}
The purpose of this common normalisation is to make a comparison 
of the coefficients $ b^{n}_{{\cal R},i}$ for different ${\cal R}$, meaningful.

The basic idea now is that the coefficients $b^{n}_{{\cal R},i}$
encode the `shape' of the state corresponding to the wavefunctional,
because they multiply the same operators, albeit in different
representations, and with a common normalisation. So to compare
our excited $k=1$ and $k=2A$ wavefunctionals we need simply
compare their coefficients  $b^{n}_{f,i}$ and  $b^{n}_{2A,i}$.
This would be straightforward if the $\phi_i$ basis operators were 
orthogonal, but since they are not, and some have very large overlaps 
with each other, apparently different sets of $b_i$'s can in fact 
correspond to very similar wavefunctionals. Normally the obvious way to 
surmount such obstacle would be to calculate directly the overlap 
of the two wavefunctionals we want to compare. Unfortunately here
that does not work because the operators have different $N$-ality
so that the overlap will be zero. So to proceed we need to transform 
the  $N$-ality of the $k=2A$ wavefunctional to $k=1$ while preserving
its essential spatial characteristics. We do so by the
simple substitution  (with $t=0$ throughout)
\begin{equation}
{{\Phi}}^n_{2A}
=
\sum^{n_o}_{i} b^{n}_{{2A},i} {\rm Tr}^{\prime}_{2A} (\phi_i)
\quad \longrightarrow \quad
{\tilde{\Phi}}^n_{2A}
=
\sum^{n_o}_{i} b^{n}_{{2A},i} {\rm Tr}^{\prime}_{f} (\phi_i)
\label{phie3}
\end{equation}
where we replace  ${\rm Tr}^{\prime}_{2A}$ by ${\rm Tr}^{\prime}_{f}$ 
in the expression for $\Phi^n_{2A}$. Intuitively it is plausible to 
think of ${\tilde{\Phi}}^n_{2A}$ as having approximately the same 
`shape' as  $\Phi^n_{2A}$, so that if we now compare
${\tilde{\Phi}}^n_{2A}$ directly with any  $\Phi^{n^\prime}_{f}$ 
by the calculation of their normalised overlap
\begin{equation}
O_{n_\prime,n}=
\frac{<{\Phi^{n^\prime\dagger}}_f {\tilde{\Phi}^n}_{2A}>}
{<{\Phi^{n^\prime\dagger}}_f \Phi^{n^\prime}_f>^{1/2} 
<{{\tilde{\Phi}}}^{n\dagger}_{2A} \tilde{\Phi}^{n}_{2A}>^{1/2} }
\label{overlap}
\end{equation}
we are in fact comparing the shape of  $\Phi^{n^\prime}_{f}$ with
that of our original ${{\Phi}}^n_{2A}$ wavefunctional.

The above method of comparison clearly has a strong heuristic
component and to some extent our confidence in the result will depend 
on how clear-cut it proves to be. So we take our four lightest
$k=2A$ (variational) eigenstates, form the corresponding set of
$\{ {\tilde{\Phi}}^n_{2A}; n=0,...,3\}$, and take the overlap
of each one of these with the corresponding $k=1$  (variational) 
eigenstates, $\{ {{\Phi}}^n_{f}; n=0,...,3\}$. Note that the
former are not necessarily orthogonal, while the latter are,
by construction (just as were the original $\{ {\Phi}^n_{2A}\}$).
We begin by showing the results, in 
Fig.~\ref{Overlap2AL32}, for our longest $l=32a$ flux tube in SU(5). 
We see that the $k=2A$ ground state has an overlap
that is  virtually $100\%$ on the $k=1$ ground state.
Since we have already established that
both ground states are to a very good approximation described
by the free string theory, this result serves to give us confidence
in this method of comparison. Turning now to the first excited 
$k=2A$ state, which is the one whose identity we are most interested 
in here, we see from Fig.~\ref{Overlap2AL32} that this has 
an overlap that is almost entirely on the first excited
$k=1$ state. Since we have established that the latter is a 
Nambu-Goto-like stringy excitation, we infer that so is the
first excited $k=2A$ state. Thus we have our answer: this state
is tending to Nambu-Goto and is not an `extra' state involving
massive excitations of the flux tube.

We also see from  Fig.~\ref{Overlap2AL32} that the next two
$P=+$, $q=0$ $k=2A$ excited states are very similar to the
corresponding $k=1$ states and therefore also Nambu-Goto-like.
In  Fig.~\ref{Overlap2AL16} we show the corresponding plots for 
length $l=16a$. The results for the ground and first excited states 
are unambiguous, but the comparison for the higher excited states is
now less clear-cut. Perhaps not surprisingly, the states become
more clearly Nambu-Goto-like as we go to longer strings.

We have repeated this whole exercise for the $k=2S$ representation.
Once again we find that the ground state has an almost $100\%$
overlap onto the $k=1$ ground state. However the first excited state
is less clear-cut than for $k=2A$, and the comparison for other states 
produces no useful information at all.

We have learned that the first excited $k=2A$ state in the $P=+,q=0$
sector is in fact Nambu-Goto-like, so it has two phonons of momentum
$p=\pm 2\pi/l$. The fact that there is such a large correction
to the free string energy when $l$ is not very large, tells us that 
these phonons have a large interaction energy.  On the other hand we
have seen that the ground state  $k=2A$ state with $P=+,q=2$
has an energy very close to the free string energy at all $l$.
This state has two phonons of the same momentum $p=2\pi/l$.
Thus we infer that the interaction energy of two phonons with 
the same momentum is very small. So although we have not found
an `extra' state, we have learned something very specific and
interesting about the inter-phonon interactions.

\section{Conclusions}
\label{section_conc}

In this paper we have examined to what extent $k=2$ flux tubes
that are wound around a spatial torus can be described by 
an effective string theory. This study complements our 
earlier study of fundamental ($k=1$) flux loops
\cite{es_k1},
where we found that a simple Nambu-Goto free string description
described the spectrum very well, even when the length of the
flux tube was on the order of its width. Since one can regard
a $k=2$ flux loop as a bound state of two $k=1$ flux loops,
one would naively expect the extra internal structure to lead
to larger corrections to any asymptotic free string description,
and indeed that there should be extra states that do not look like 
excited string states and which arise through the excitation of 
the massive binding modes. 

Because $k=2$ states are much more massive than $k=1$ states, some
systematic errors are much more important for the former. This has 
meant that our most reliable calculations are for the ground states 
in the various sectors labeled by different quantum numbers. These
quantum numbers are the parity $P$ and the momentum along the flux
tube, $q$. Since we expect the ground state unexcited flux loop
to be invariant under translations along its axis, and to be symmetric 
under reflections across its axis, having $P\neq +$ and/or $q\neq 0$
explicitly probes the non-trivial excitation spectrum of the flux loop
(in a way that transverse momenta do not, which is why we do not
consider them).

In addition to these exact quantum numbers, we have confirmed 
by explicit calculation that there is very little overlap between the totally 
symmetric ($k=2S$) and totally antisymmetric ($k=2A$) representations.
We thus find that our variational eigenfunctionals fall into these
representations to a good approximation, so that we can categorise
the spectra by their behaviour under the full SU($N$) group, and not just
its center $Z_N$. 

In the $k=2A$ sector we found that the ground states for $q=0,1,2$ have
energies that are very close to those of the free bosonic string theory
(Nambu-Goto in flat $D=2+1$). The leading permitted correction to 
$\left(E(q=0;l)/(\sigma l)\right)^2$ is $O(1/(\sqrt{\sigma}l)^5)$, and its fitted
coefficient has a natural value of $O(1)$. This is in contrast to 
the fundamental flux loop where the corresponding coefficient
is unnaturally tiny, $<O(1/10)$. More surprising is that the
$q = 1,2$ ground state energies are well described by Nambu-Goto
all the way down to $l\sqrt{\sigma} \simeq 1.5$. Such a flux tube is
a blob rather than a string and it is remarkable that the thin string
oscillation spectrum should accurately describe its lowest modes.
Just as for the $k=1$ flux tube, one is tempted to read into this
evidence for some kind of gauge-string duality that makes a thin
string the starting point for describing even a short blob-like 
flux tube. For the low-lying excited states in the $q=0$ sector the 
deviations are much larger -- and of a size that one would expect within an
effective string theory description that was only valid asymptotically.

The much heavier $k=2S$ sector is clearly afflicted by much larger
systematic errors, which strongly limits our ability to draw any
useful conclusions, although here too the ground states of various
momenta $q$ are recognisably string like.  

As an interesting aside, by including operators that wound twice
around the torus, but such that the second winding overlapped little 
in space with the first, we were able to identify some eigenstates that 
correspond to doubly wound fundamental flux loops, in addition to the 
bound $k=2$ singly wound flux loops. 

While it is very informative to observe how well the various
states are described by simple free string excitations, this can
only be the first step. We also need to pin-point the nature of the 
corrections to the free string theory. This involves not only quantifying 
and searching for patterns in the corrections to the free string 
excitations -- the interactions amongst the phonons -- but also
identifying excited states that are additional to those associated
with simple string oscillations. Such additional states, due to excitations 
of massive modes, would normally be predicted by any theoretical approach.

For example, in a naive field theoretic approach, where the flux tubes 
are some kind of dynamically generated non-Abelian Nielsen-Olesen vortices,
there should be massive excitations associated with the nontrivial flux 
tube structure. In an `AdS/CFT' approach there would be a highly curved
metric in the region of the higher dimensional space 
where the string picks up its non-trivial energy per unit length. 
Naively, the energy of such an `extra' state should be some (roughly) 
constant amount $\mu$ above the ground state energy. As we remarked earlier, 
a naive estimate for $\mu$ might be the lightest glueball mass,
$\sim 4\surd\sigma_f$, or possibly the gap between the lightest glueball 
masses, $\sim 2\surd\sigma_f$. On such an estimate, such states should be 
easily visible for small flux tube lengths, where they would be lighter 
than some of the simple thin string excitations. A glance at the $q=0$ 
spectra in Figs.~\ref{spectrum2AP+SU4} and \ref{spectrum2AP+SU5}, 
immediately suggests that the first excited state could be a candidate
for such a state. However our analysis in Section~\ref{section_extra}
of the wavefunctional of that
state showed that its `shape' was too much like that of the corresponding
$k=1$ flux tube, which is unambiguously Nambu-Goto-like, to believe
that it is anything other than a simple string excited state. We have in
fact not observed any states, in the range of energies where we have some
control, that could be interpreted as such non-stringy excitations.
This could either mean that we have a poor overlap onto such states
or that their masses are much larger than expected. (The glueball
mass might not be a good guide if its dynamics is that of a string-like 
contractible closed loop of flux, so that neither its mass nor the
gap to the first excited state are strongly influenced
by the dynamics associated with the non-stringy massive modes.)
In the former case it means
that we must suitably extend our basis of operators; and for that purpose
it would be useful to have some guidance from, say, the 
gauge-gravity side, about the lightest such extra states.
In the latter case, the implication would be that the non-stringy
physics was essentially decoupled from the low-energy physics of the
confining gauge theory. Resolving this issue would provide something
very interesting in either case.

Although the question of non-stringy `extra' states has not been settled
in this paper, we have been able to identify some qualitative features
of the interactions amongst the phonons along the flux tube. As we have
just remarked, we have shown that the first excited $q=0$ $P=+$ state 
is in fact a string excited state, i.e. it has two phonons of opposite 
momentum $p=\pm 2\pi/l$. And, as we see  in Figs.~\ref{spectrum2AP+SU4} 
and \ref{spectrum2AP+SU5}, the energy of 
this state has very large corrections to the corresponding free 
string energy. On the other hand the lightest $q=2$  $P=+$ state, which
has exactly the same two phonons, but with the same momentum, has
as we see in  Figs.~\ref{SU42Ars} and ~\ref{SU52Ars}, an energy that
is virtually identical to that of the  corresponding free 
string state, over our whole range of $l$. Thus we can come to a
very specific conclusion about the non-trivial interactions 
between two phonons: they are negligible when the phonons have exactly 
the same momenta (and are therefore at `threshold'), 
but are large when the momenta are exactly opposite. 

One might ask why such a conclusion had not already been reached in our
much more accurate calculations of the fundamental $k=1$ flux loop
spectrum in
\cite{es_k1}.
There are at least two plausible reasons.
Firstly, we have seen that the corrections are 
in general much smaller for $k=1$ than for $k=2$ flux loops, so much
so that the first excited 
$q=0$ $P=+$ $k=1$ flux loop looks, at a first glance, to be close 
to the free string prediction. Secondly, we observed in
\cite{es_k1}
some significant corrections in the $q=2$ spectrum at small $l$,
in contrast to the case of $k=2$. 
However we have since been able to show that these 
are largely due to lattice spacing corrections to the energy-momentum  
dispersion relation. Thus a renewed analysis may well show that the $k=1$
flux loop displays the same kind of non-trivial interaction between
phonons that we have seen with the $k=2$ loop. 
The fact that the $k=2$ calculation in this
paper has been performed at a smaller value of $a$,
together with the fact that the $k=2$ loops are much heavier 
and therefore less sensitive to the momentum and to any lattice
corrections thereof, may be part of the reason that 
it has been easier to identify 
this effect in the present, nominally more difficult, calculation.

There is of course much more information about the phonon interactions,
that is implicit in our calculations. We have focused on the above 
single result because it involves energy levels that are non-degenerate,
making the argument and conclusion particularly straightforward. In most
other cases corrections to the Nambu-Goto energy levels are accompanied
by splittings and (presumably) mixing of the would-be degenerate string states,
and this will complicate the analysis. While there are other examples
of non-degenerate states that can be usefully compared, e.g. the ground 
state in the $P=+, q=3$ sector and the first excited state in the
$P=+, q=1$ sector, these involves values of momentum $q > 2$ which
have not been studied in the $k=2$ calculation of the present paper
(but have been included in our accompanying $k=1$ calculations).
We therefore leave a more complete analysis to a
forthcoming publication
\cite{all_k}
that will describe in detail all our $k=1$ and $k=2$ calculations.

\section*{Acknowledgements}

MT thanks the Galileo Galilei Institute for its hospitality
during the Workshop on Non-Perturbative Methods in Strongly Coupled
Gauge Theories and various participants throughout the workshop 
for very useful and informative discussions.
The computations were performed on resources funded by Oxford and EPSRC. 
AA was supported by the EC 6$^{th}$ Framework Programme Research and 
Training Network MRTN-CT-2004-503369 and the Leventis Foundation. 
BB was supported in part by the U.S. Department of Energy under 
Grant No. DE-FG02-96ER40956.

\newpage
\section*{Appendix A: Construction of operators}

To calculate the excitation spectrum of $k=2$ strings, it is 
necessary to find a way to project onto such states. The way to achieve 
this is to find a suitable basis described by good quantum numbers, 
in which our operators will be encoded. In our case, this basis is 
defined by the quantum numbers of momentum, $q$, along the string, and the parity 
$P$, where the latter tells us how a string transforms under reflections in
its transverse axis. This imples that we need to introduce transverse 
deformations in Polyakov loops and construct line paths that transform 
in a certain way under such parity reflections and longtitudinal momentum. In general, the more operators we 
use the better the results we obtain; we therefore construct a 
plethora of Polyakov paths trying to extract states with high overlaps. 

In the first step of the calculation we construct the most commonly used $k=2$ string 
operators: $\left({\rm Tr}\{ l_p \}\right)^2$ and ${\rm Tr}\{ l^2_p \}$. Our desire 
to extract the excitation spectrum and introduce more degrees of freedom, 
such as the transverse parity and momentum, forces us to modify them 
and thus to complicate their structure, as discussed 
in Section \ref{subsection_loops}. In Eq.~(\ref{k1}-\ref{k21}) we 
demonstrate how these new operators are composed for the $k=1,2$ cases. For simplicity, below we demonstrate  the way our operators have been constructed with a particular transverse deformation that is easy to visualize. First are the negative and positive parity operators for the $k=1$ strings:
\begin{eqnarray}
\Phi^{P=\pm}_{k=1}=\rm{Tr} \ \left\lbrace \ \parbox{1cm}{\rotatebox{0}
{\includegraphics[width=1cm]{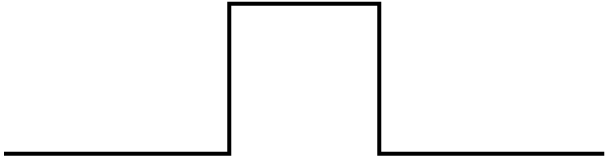}}} \ \right\rbrace \pm \rm{Tr}  
\left\lbrace \ \parbox{1cm}{\rotatebox{180}
{\includegraphics[width=1cm]{square1w.eps}}} \ \right\rbrace. \label{k1}
\end{eqnarray}
From these we construct the two simplest sets of positive and negative parity operators for the $k=2$ string, i.e. those corresponding to ${\rm Tr} \, \{l_p \}^2$
\begin{eqnarray}
\Phi^{P=\pm}_{k=2}=\rm{Tr} \ \left\lbrace \ \parbox{1cm}{\rotatebox{0}{
\includegraphics[width=1cm]{square1w.eps}}} \ \right\rbrace^2 \pm \rm{Tr}  
\left\lbrace \ \parbox{1cm}{\rotatebox{180}
{\includegraphics[width=1cm]{square1w.eps}}} \ \right\rbrace^2,
\label{k2}
\end{eqnarray}
and those corresponding to  ${\rm Tr} \, \{l_p ^2\}$
\begin{eqnarray}
\Phi^{P=\pm}_{k=2}=\rm{Tr} \ \left\lbrace \ \parbox{1cm}{\rotatebox{0}
{\includegraphics[width=1cm]{square1w.eps}}} \cdot \parbox{1cm}
{\rotatebox{0}
{\includegraphics[width=1cm]{square1w.eps}}}  \ 
\right\rbrace \pm \rm{Tr}  \left\lbrace \ \parbox{1cm}
{\rotatebox{180}
{\includegraphics[width=1cm]{square1w.eps}}} 
\cdot  \parbox{1cm}{\rotatebox{180}
{\includegraphics[width=1cm]{square1w.eps}}}  \ \right\rbrace.
\label{k21}
\end{eqnarray}

Next, we project onto the $k=2$ totally antisymmetric and 
symmetric representations. To single-out the irreducible representations 
which describe the theory, we perform antisymmetrisation and symmetrisation 
according to the relevant Young-tableau decomposition of $k=2$ fundamental 
colour sources: 
$\parbox{0.3cm}{\rotatebox{270}
{\includegraphics[width=0.3cm]{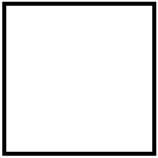}}} 
\bigotimes  \parbox{0.3cm}{\rotatebox{270}
{\includegraphics[width=0.3cm]{square.eps}}} =  \parbox{0.6cm}{\rotatebox{0}
{\includegraphics[width=0.6cm]{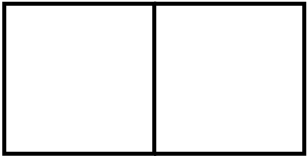}}} 
\bigoplus  \parbox{0.3cm}{\rotatebox{90}
{\includegraphics[width=0.6cm]{rectangular.eps}}} $. 
The resulting operators will be of the type: 
${\rm Tr}\{ l_p \}^2 \pm {\rm Tr}\{ l^2_p \}$, where $+(-)$ for 
symmetric(antisymmetric) representation. Once more we need to introduce 
transverse deformations, in order to project onto the non-trivial 
irreducible representations that characterise the closed flux tube in 
$D=2+1$. Examples of such operators are demonstrated in 
Eqs.~(\ref{symm}, \ref{antsymm}) below. Begining with the projection onto the $k=2$ antisymmetric representation we find
\begin{eqnarray}
\Phi^{P=\pm}_{k=2A}= [ \rm{Tr} \ \left\lbrace \ \parbox{1cm}{\rotatebox{0}
{\includegraphics[width=1cm]{square1w.eps}}}
 \ \right\rbrace^2 - \rm{Tr} \ \left\lbrace \ \parbox{1cm}{\rotatebox{0}
{\includegraphics[width=1cm]{square1w.eps}}} \cdot \parbox{1cm}{\rotatebox{0}
{\includegraphics[width=1cm]{square1w.eps}}}  \ 
\right\rbrace ] \pm [ \rm{Tr}  \left\lbrace \ 
\parbox{1cm}{\rotatebox{180}
{\includegraphics[width=1cm]{square1w.eps}}} \ \right\rbrace^2 - \rm{Tr}  
\left\lbrace \ \parbox{1cm}{\rotatebox{180}
{\includegraphics[width=1cm]{square1w.eps}}} \cdot  \parbox{1cm}{\rotatebox{180}
{\includegraphics[width=1cm]{square1w.eps}}}  \ \right\rbrace],
\label{symm}
\end{eqnarray}
and projecting onto the $k=2$ symmetric representation one obtains
\begin{eqnarray}
\Phi^{P=\pm}_{k=2S}= [ \rm{Tr} \ \left\lbrace \ \parbox{1cm}{\rotatebox{0}
{\includegraphics[width=1cm]{square1w.eps}}} \ 
\right\rbrace^2 + \rm{Tr} \ \left\lbrace \ \parbox{1cm}{\rotatebox{0}
{\includegraphics[width=1cm]{square1w.eps}}} \cdot 
\parbox{1cm}{\rotatebox{0}
{\includegraphics[width=1cm]{square1w.eps}}}  \ 
\right\rbrace ] \pm [ \rm{Tr}  \left\lbrace \ \parbox{1cm}{\rotatebox{180}
{\includegraphics[width=1cm]{square1w.eps}}} \ 
\right\rbrace^2 + \rm{Tr}  \left\lbrace \ \parbox{1cm}{\rotatebox{180}
{\includegraphics[width=1cm]{square1w.eps}}} \cdot  
\parbox{1cm}{\rotatebox{180}
{\includegraphics[width=1cm]{square1w.eps}}}  \ \right\rbrace].
\label{antsymm}
\end{eqnarray}
As the notation suggests, in both cases above the $\pm$ signs determine the parity $P=\pm$ of $\Phi$. The complete set of polyakov lines used in our calculation is presented 
in Table~\ref{Operators}.

As mentioned in Section~\ref{subsection_loops}, we have also attempted 
to test whether the $k=2$ string spectrum includes unbound $w=2$ states,
which we expect will only appear in our calculation
if we use the appropriate operators. These new states 
are expected to be described by frequencies lower than those describing 
the $w=2$ bound states. Although some $k=2$ operators look as if they wind 
twice around the torus i.e Eqs.~(\ref{k2}--\ref{k21}), the way we had been constructing 
them prohibits us to project onto states with lower frequencies since each 
Polyakov loop starts and ends at the same lattice point within one lattice 
size. To overcome this we construct Polyakov lines that wind twice around the 
torus with transverse deformations at the joint of the two lattices as in : 
\begin{eqnarray}
\Phi^{P=\pm}_{w=2}=\rm{Tr} \ \left\lbrace \ \parbox{2cm}{\rotatebox{0}
{\includegraphics[width=2cm]{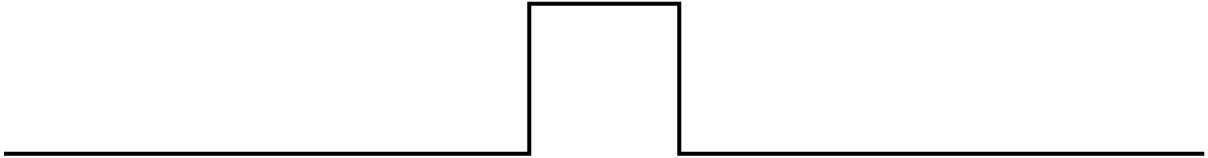}}} \ 
\right\rbrace \pm \rm{Tr}  \left\lbrace \ \parbox{2cm}{\rotatebox{180}
{\includegraphics[width=2cm]{square2w.eps}}} \ \right\rbrace.
\label{w2}
\end{eqnarray}
Note the number of lattice links in the $x$ directions that goes into each of the operators in the traces above is $2L_x$, i.e. twice the lattice size. This is in contrast to the case of Eqs.~(\ref{k2})--(\ref{k21}), where it is only $L_x$. 

Finally, let us note that the new operators in Eq.~\ref{w2} transform 
the same under the centre of the group $Z_N$ as the other $k=2$ operators 
i.e $\Phi_{w=2}={\rm Tr} (l_p {l'}_p) \to z^2 {\rm Tr} (l_p {l'}_p)$. So 
these new `unbound' operators have $N$-ality $k=2$ and will therefore contribute to 
an extended $k=2$ spectrum.

\vfil\eject

\vfil\eject

\begin{table}
\begin{center}
\begin{tabular}{||c||c||c||c||c|c||} \hline \hline
\ level \ & \ $N_R$ \ & \ $N_L$ \ & \ $q$ \ & \ \ $P=+$  \ \ & \ \ 
$P=-$ \ \ \\  \hline
$0$   &  $0$  &  $0$  & $0$ & $|0\rangle$ & $$ \\ \hline
$1$   & $1$   &  $0$  & $1$ & $$ & $\alpha_{-1}|0\rangle$   \\ \hline
$2$   &  $1$  &  $1$  & $0$ & $\alpha_{-1} \bar{\alpha}_{-1}|0\rangle$ & 
$$ \\ \hline
$3$   & $2$   &  $0$  & $2$ & $\alpha_{-1} \alpha_{-1}|0\rangle$ & 
$\alpha_{-2}|0\rangle$  \\ \hline
$4$   & $2$   &  $1$  & $1$ & $\alpha_{-2} \bar{\alpha}_{-1}|0\rangle$ & 
$\alpha_{-1} \alpha_{-1}  \bar{\alpha}_{-1}|0\rangle$  \\ \hline
$5$   &  $2$  &  $2$  & $0$ & $\alpha_{-2} \bar{\alpha}_{-2}|0\rangle$,  
\ $\alpha_{-1}\alpha_{-1} \bar{\alpha}_{-1} \bar{\alpha}_{-1} |0\rangle$  
& $\alpha_{-2} \bar{\alpha}_{-1} \bar{\alpha}_{-1}|0\rangle$, \ 
$\alpha_{-1}\alpha_{-1} \bar{\alpha}_{-2}|0\rangle$  \\ \hline
$6$   & $3$   &  $1$  & $2$ & $\alpha_{-3} \bar{\alpha}_{-1}|0\rangle$,  
\ $\alpha_{-1} \alpha_{-1} \alpha_{-1}  \bar{\alpha}_{-1}|0\rangle$ & 
\ $\alpha_{-2} \alpha_{-1}  \bar{\alpha}_{-1}|0\rangle$ \\ \hline \hline 
\end{tabular} 
\caption{The seven lowest Nambu-Goto energy levels for the $w=1$ closed 
string. If the number of creation operators is even(odd) then the 
state has positive(negative) parity.}
\label{levels}
\end{center}
\end{table}

\begin{table}
\begin{center}
\begin{tabular}{|c|c||c|c|c||c|c|}\hline
\multicolumn{7}{|c|}{ $O_{AS}(t=0)$ }  \\ \hline 
$b_A$ &  $b_S$  &  $l=l_y=16$ &   $l=16,l_y=20$ & 
 $l=16,l_y=48$ & $l=l_y=20$ & $l=l_y=24$  \\ \hline 
1 & 1 &  0.0002(3) &  0.0001(13) & -0.0014(13) & -0.0001(3)  & -0.0001(4)   \\
2 & 2 & -0.0001(3) &  0.0020(10) & -0.0010(9)  &  0.0001(2)  & -0.0001(3)   \\
3 & 3 & -0.0001(3) & -0.0004(14) & -0.0003(10) & -0.0001(3)  &  0.0003(4)   \\
4 & 4 &  0.0025(3) &  0.0001(18) & -0.0017(12) &  0.0003(4)  & -0.0002(4)   \\
4 & 5 &  0.1822(3) &  0.0589(19) &  0.0277(10) &  0.0142(4)  & -0.0027(4)   \\
5 & 4 &  0.0353(3) &  0.0118(16) &  0.0007(13) &  0.0040(3)  &  0.0003(3)   \\
5 & 5 &  0.2806(4) &  0.1335(22) &  0.0404(13) &  0.0472(4)  &  0.0005(5)   \\ \hline
\end{tabular}
\caption{\label{table_OAS}
Overlaps of Polyakov loops in the k=2A and k=2S representations,
with blocking levels $b_A$ and $b_S$ respectively, at $t=0$ and 
as defined by eqn(\ref{eqn_OAS}).
For SU(4) at $\beta=32.0$ on $l\times l_y\times 20$ lattices.}
\end{center}
\end{table}

\begin{table}
\begin{center}
\begin{tabular}{|c|c||c|c|c|}\hline
\multicolumn{5}{|c|}{ $O_{AS}(t) \quad 24^2 32$ }  \\ \hline 
$b_A$ &  $b_S$  &  $t=0$ &   $t=1$ &  $t=2$ \\ \hline 
1 & 1 &  0.0001(2) &  0.0075(31) &  0.0055(186) \\
2 & 2 & -0.0000(3) &  0.0000(10) & -0.0010(27)  \\
3 & 3 &  0.0002(3) &  0.0002(6)  & -0.0017(14) \\
4 & 4 &  0.0008(4) &  0.0010(7)  &  0.0017(12) \\
4 & 5 &  0.0098(5) &  0.0151(8)  &  0.0224(13) \\
5 & 4 &  0.0043(4) &  0.0059(7)  &  0.0086(13) \\
5 & 5 &  0.0292(5) &  0.0380(9)  &  0.0489(14) \\ \hline
\end{tabular}
\caption{\label{table_OASt}
Overlaps of Polyakov loops in the k=2A and k=2S representations,
with blocking levels $b_A$ and $b_S$ respectively, at $t=0,1,2$ and 
as defined by eqn(\ref{eqn_OAS}).
For SU(5) at $\beta=80.0$ on $24^2 32$ lattices.}
\end{center}
\end{table}

\begin{table}
\scriptsize
\centering{
\begin{tabular}[c]{||c||c||c||c||c||c||c||c||} \hline \hline
1 & 2 & 3 & 4 & 5 & 6 & 7 & 8 \\ \hline \hline 
\includegraphics[height=3cm]{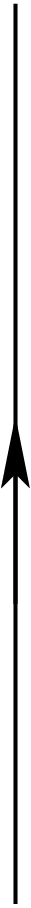} 
&
\includegraphics[height=3cm]{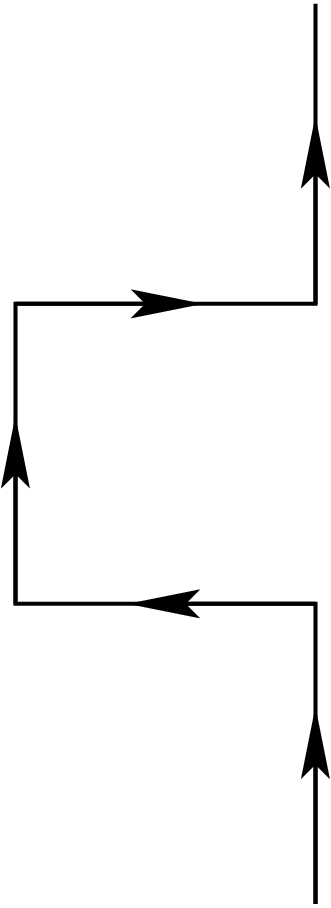}
&
\includegraphics[height=3cm]{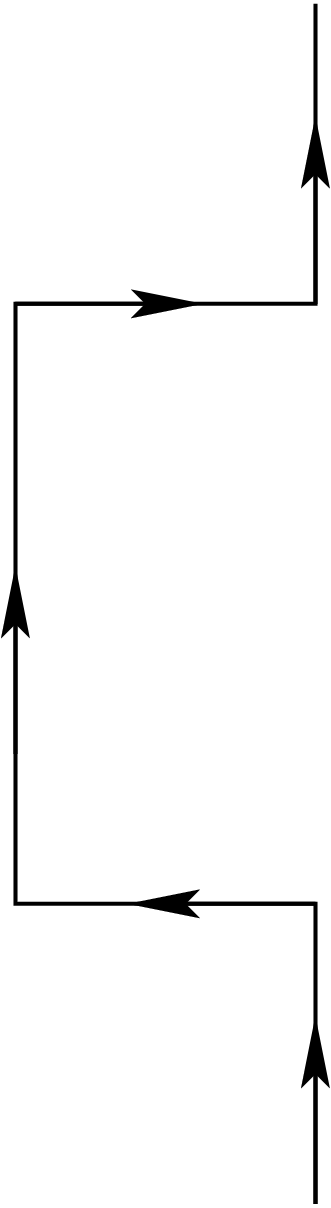}
&
\includegraphics[height=3cm]{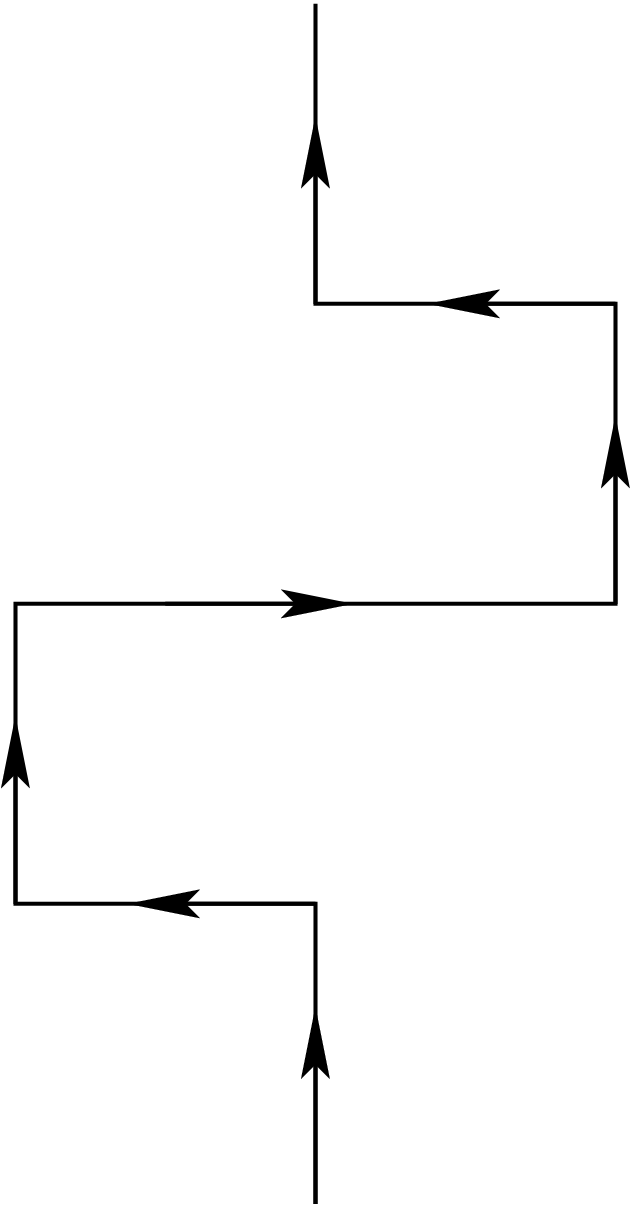} 
&
\includegraphics[height=3cm]{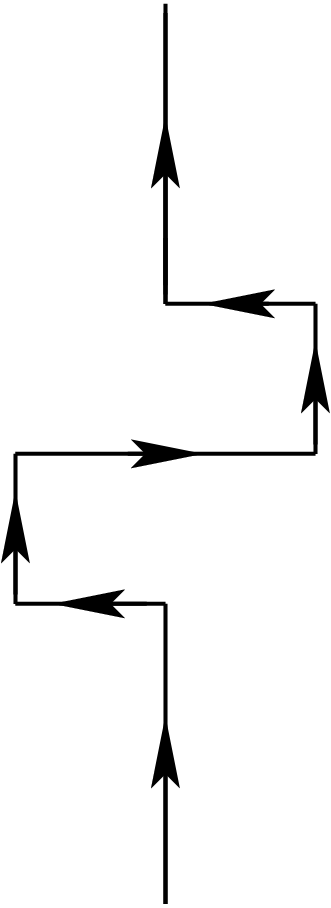} 
&
\includegraphics[height=3cm]{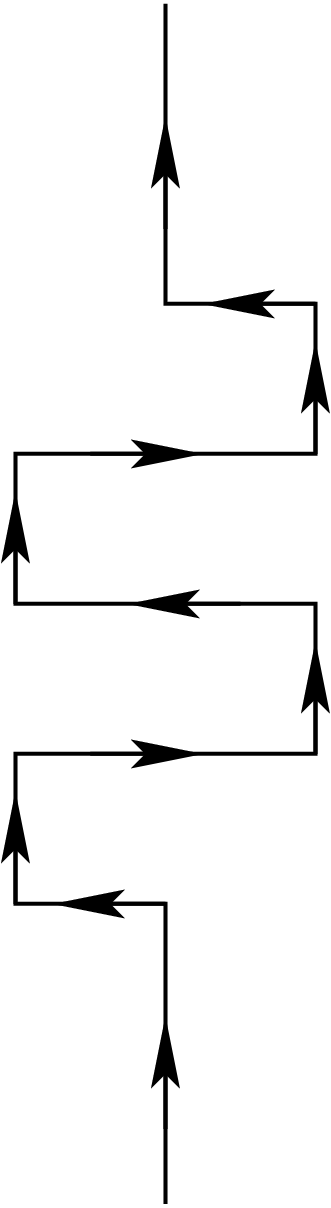} 
&
\includegraphics[height=3cm]{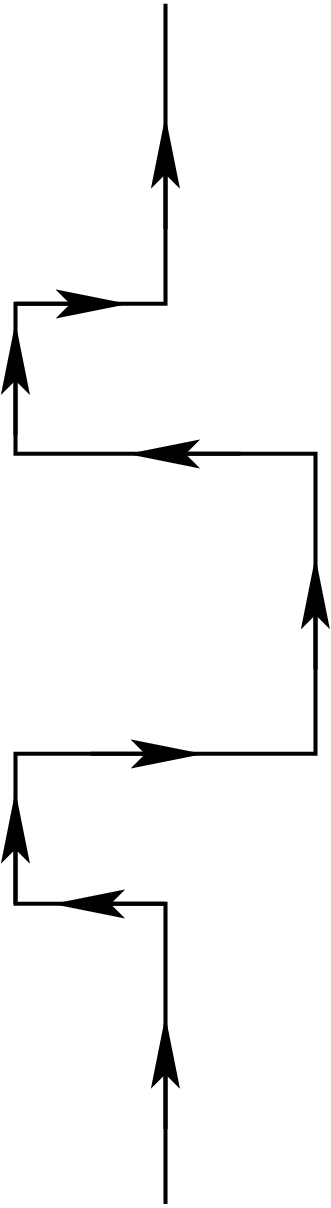} 
&
\includegraphics[height=3cm]{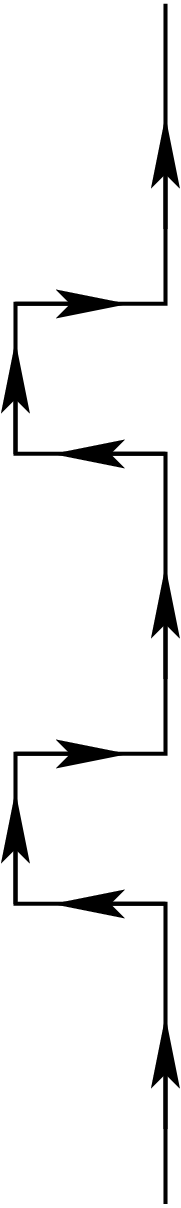} \\ \hline \hline 
9&10&11&12&13&14&15&16 \\ \hline\hline
\includegraphics[height=3cm]{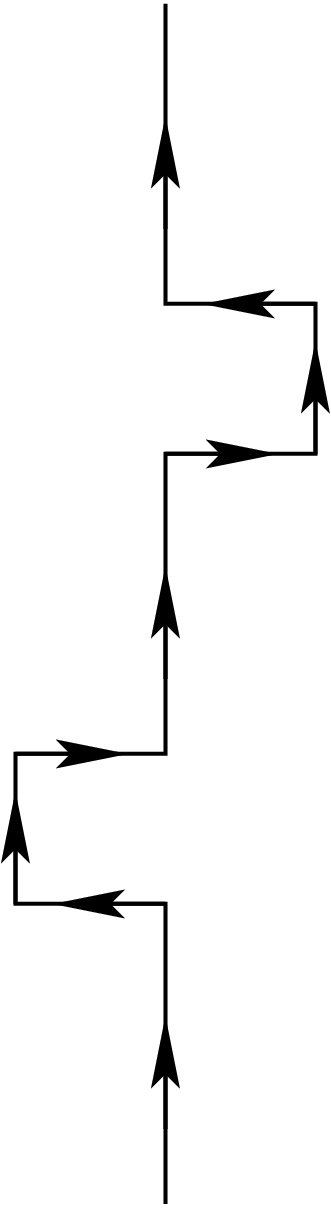} 
&
\includegraphics[height=3cm]{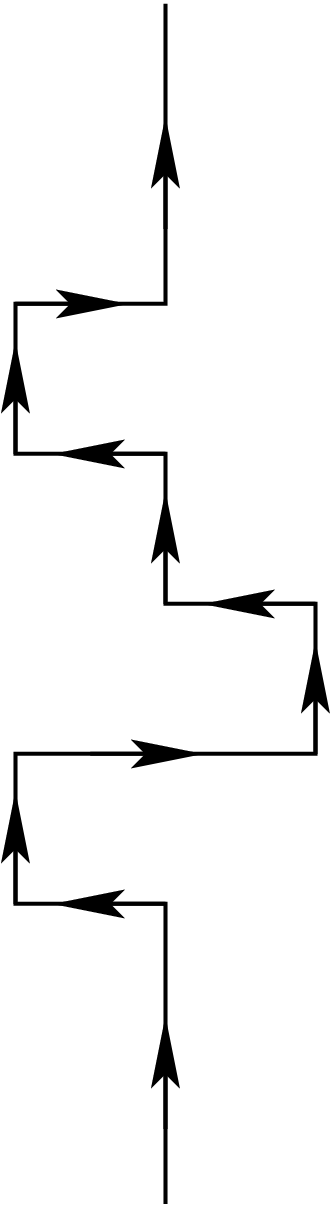} 
&
\includegraphics[height=3cm]{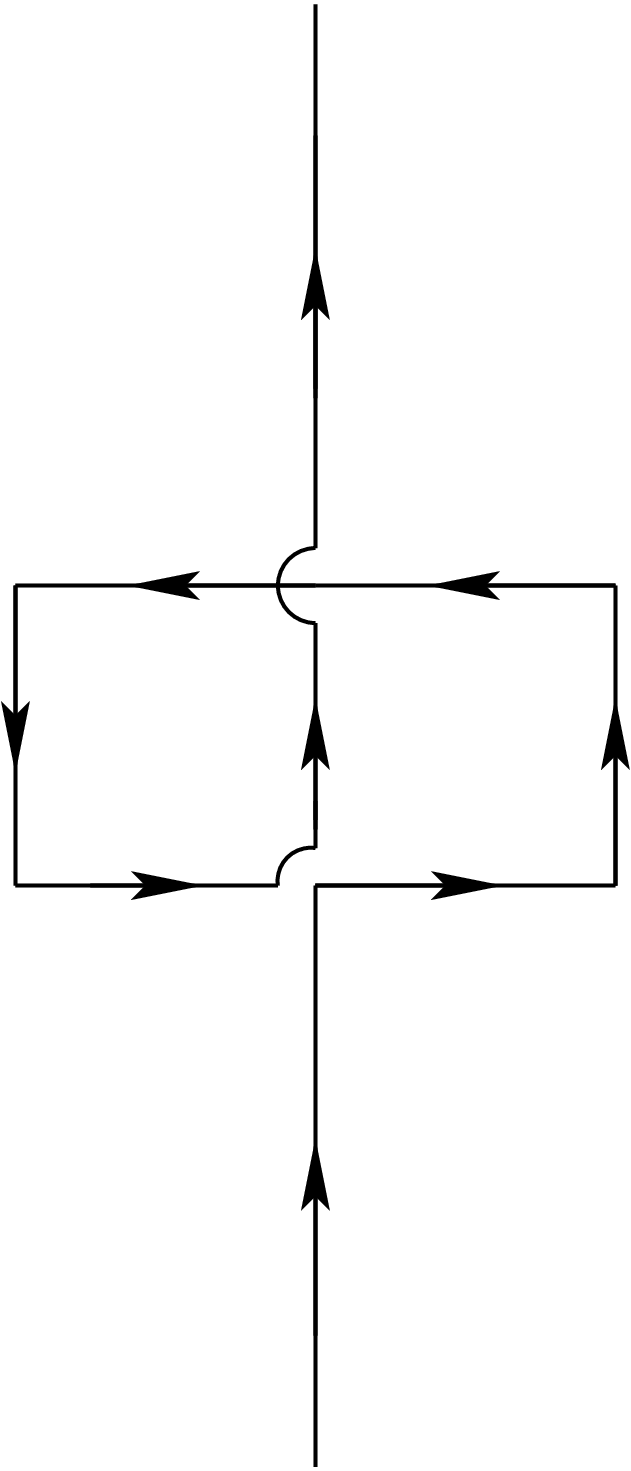} 
&
\includegraphics[height=3cm]{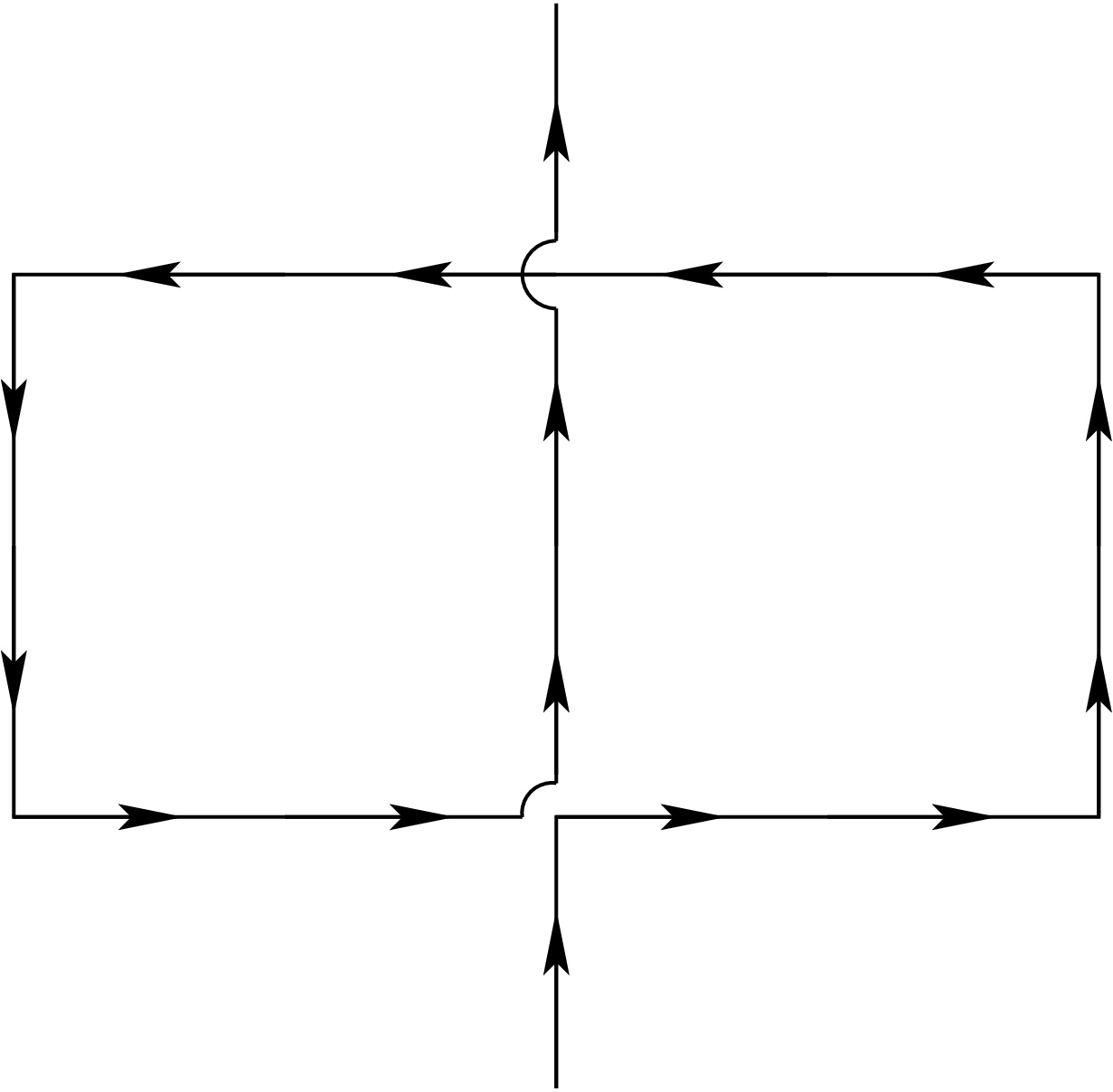} 
&
\includegraphics[height=3cm]{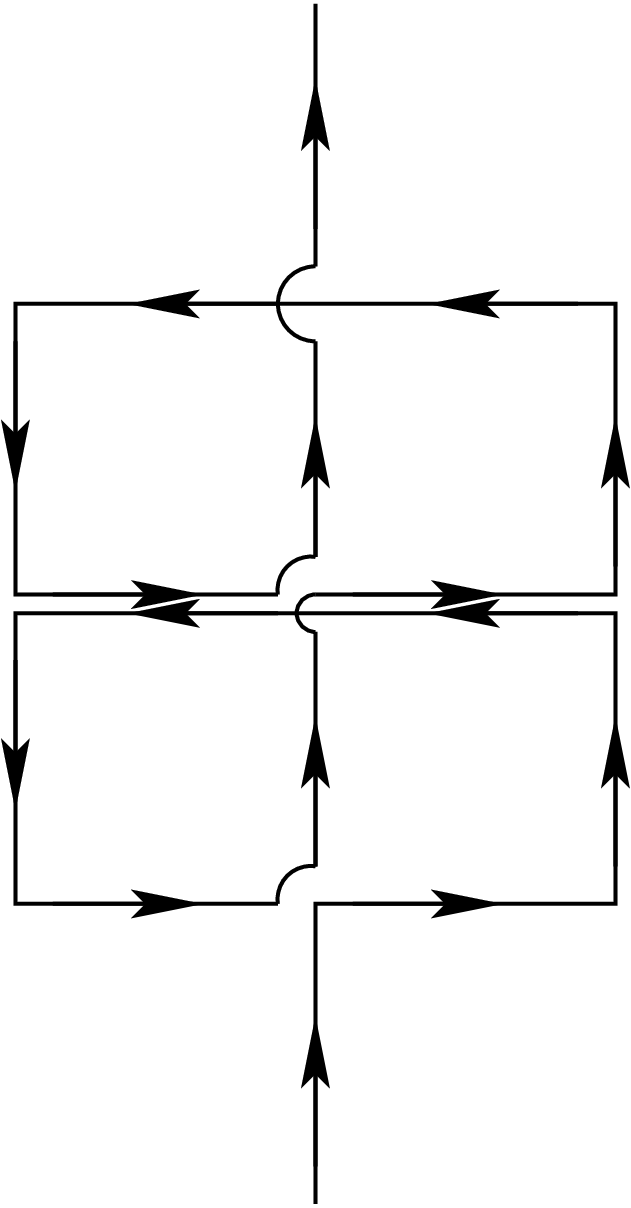}
&
\includegraphics[height=3cm]{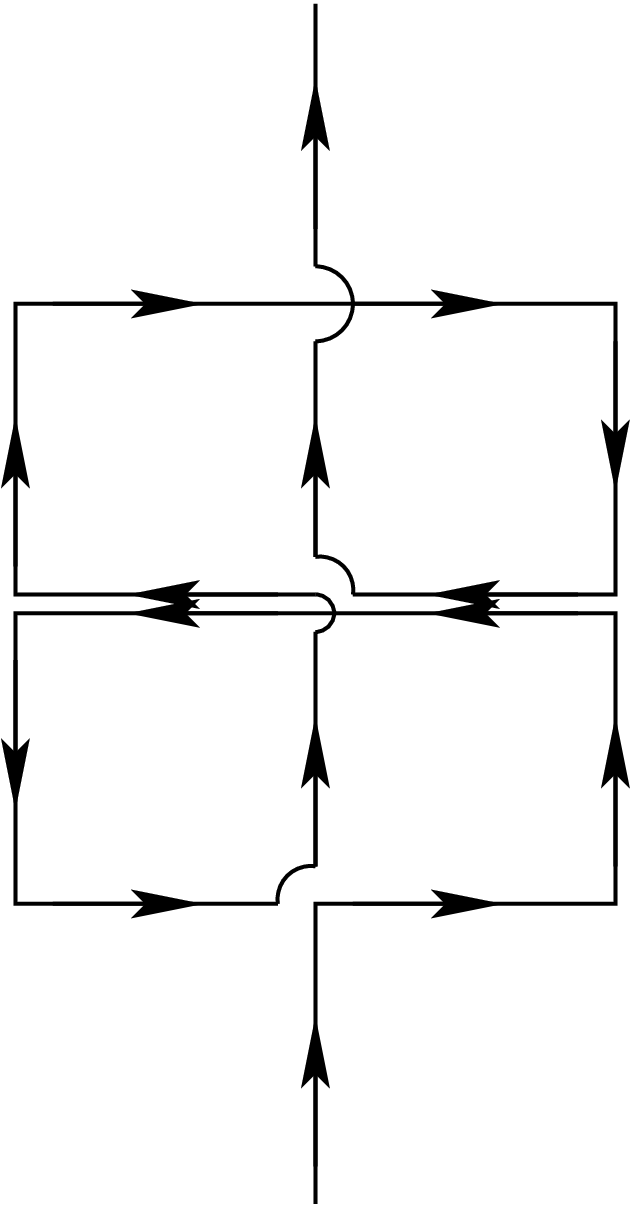}
&
\includegraphics[height=3cm]{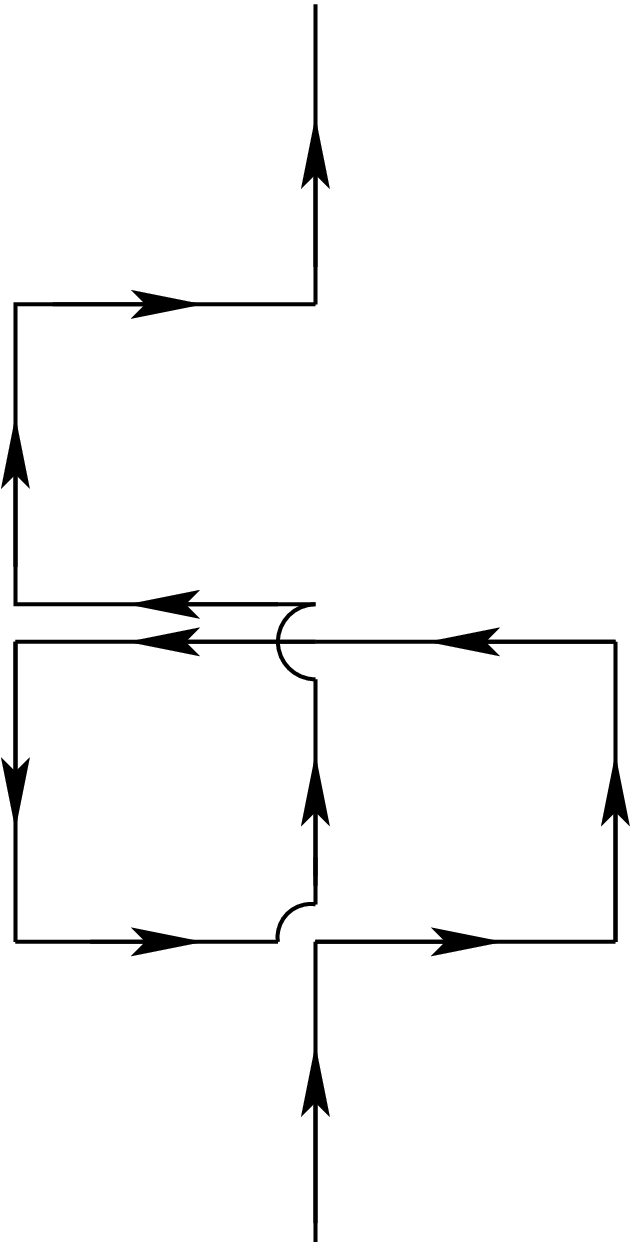}
&
\includegraphics[height=3cm]{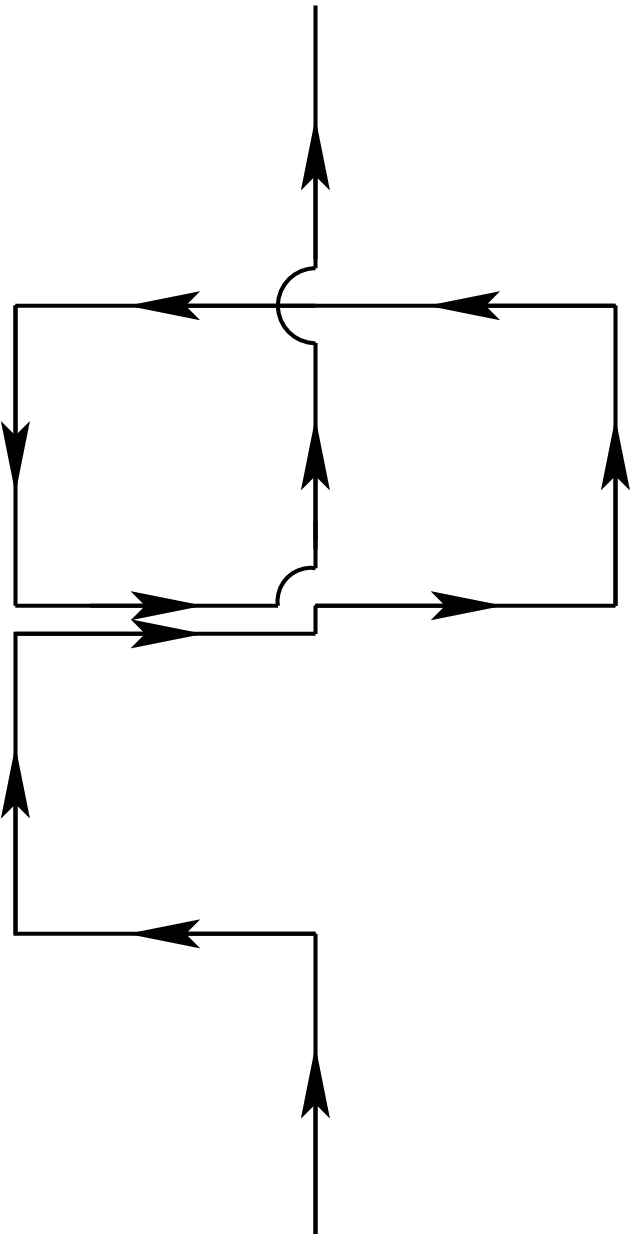} \\ \hline \hline


\end{tabular}
}
\caption{The sixteen transverse deformations used in the construction 
of the operators in this work. Each line comes in five different 
blocking/smearing levels, in 
order to enhance the overlap onto the physical states.}
\label{Operators}
\end{table}

\begin{figure}[p]
\centerline{
\scalebox{0.9}{\input{plot2ArsSU4.tex} }}
\caption{Energies of the lightest four states with $q=0,1,2$, for 
winding flux tubes in the $k=2$ antisymmetric representation of $SU(4)$ 
at $\beta=50$. Lines are the Nambu-Goto predictions of eqn(\ref{eqn_NG}), 
with the string tension obtained by fitting the ground state.}
\label{SU42Ars}
\end{figure}

\begin{figure}[p]
\centerline{
\scalebox{0.9}{\input{plot2ArsSU5.tex} }}
\caption{Energies of the lightest four states with $q=0,1,2$, for 
winding flux tubes in the $k=2$ antisymmetric representation of $SU(5)$ 
at $\beta=80$. Lines are the Nambu-Goto predictions of eqn(\ref{eqn_NG}), 
with the string tension obtained by fitting the ground state.}
\label{SU52Ars}
\end{figure}

\begin{figure}[p]
\begin{center}
\scalebox{0.9}{\input{newplotsub2ASU4.tex} }
\end{center}
\caption{Flux tube excitation energies from the SU(4) $q=1,2$ calculations
in Fig.~\ref{SU42Ars}, extracted using eqn(\ref{eqn_exq}). Lines are 
the Nambu-Goto predictions from eqn(\ref{eqn_NG}).}
\label{SU4sub2A}
\end{figure}

\begin{figure}[p]
\begin{center}
\scalebox{0.9}{\input{newplotsub2ASU5.tex} }
\end{center}
\caption{Flux tube excitation energies from the SU(5) $q=1,2$ calculations
in Fig.~\ref{SU52Ars}, extracted using eqn(\ref{eqn_exq}). Lines are 
the Nambu-Goto predictions from eqn(\ref{eqn_NG}).}
\label{SU5sub2A}
\end{figure}

\begin{figure}
\centerline{\scalebox{0.9}{\input{plotspectrum2ArsSU4.tex} }}
\caption{Energies of the lightest four states with $P=+$ and $q=0$, 
for winding flux tubes in the $k=2$ antisymmetric representation 
of $SU(4)$ at $\beta=50$. Lines are the Nambu-Goto predictions 
of eqn(\ref{eqn_NG}), 
with the string tension obtained by fitting the ground state.}
\label{spectrum2AP+SU4}
\end{figure}

\begin{figure}
\centerline{\scalebox{0.9}{\input{plotspectrum2ArsSU5.tex} }}
\caption{Energies of the lightest four states with $P=+$ and $q=0$, for 
winding flux tubes in the $k=2$ antisymmetric representation of $SU(5)$ 
at $\beta=80$. Lines are the Nambu-Goto predictions of eqn(\ref{eqn_NG}), 
with the string tension obtained by fitting the ground state.}
\label{spectrum2AP+SU5}
\end{figure}

\begin{figure}[p]
\centerline{
\scalebox{0.9}{\input{plot2SrsSU4.tex} }}
\caption{Energies of the lightest four states with $q=0,1,2$, for 
winding flux tubes in the $k=2$ symmetric representation of $SU(4)$ 
at $\beta=50$. Lines are the Nambu-Goto predictions of eqn(\ref{eqn_NG}), 
with the string tension obtained by fitting the ground state.}
\label{SU42Srs}
\end{figure}

\begin{figure}[p]
\centerline{
\scalebox{0.9}{\input{plot2SrsSU5.tex} }}
\caption{Energies of the lightest four states with $q=0,1,2$, for 
winding flux tubes in the $k=2$ symmetric representation of $SU(5)$ 
at $\beta=80$. Lines are the Nambu-Goto predictions of eqn(\ref{eqn_NG}), 
with the string tension obtained by fitting the ground state.}
\label{SU52Srs}
\end{figure}

\begin{figure}[p]
\begin{center}
\scalebox{0.9}{\input{newplotsub2SSU5.tex} }
\end{center}
\caption{Flux tube excitation energies from the SU(5) $q=1,2$ calculations
in Fig.~\ref{SU52Srs}, extracted using eqn(\ref{eqn_exq}). Lines are 
the Nambu-Goto predictions from eqn(\ref{eqn_NG}).}
\label{SU5sub2S}
\end{figure}

\clearpage

\begin{figure}[!htb]
\centerline{
\scalebox{1.0}{\input{plot} }
\put(-338,245){\scriptsize Ground State}
\put(-250,245){\scriptsize $\rm 1^{st}$ excited state}
\put(-160,245){\scriptsize $\rm 2^{nd}$ excited state}
\put(-70,245){\scriptsize $\rm 3^{rd}$ excited state}
\put(-375,200){$|O_{n_{\cal F},0}|^2$}
\put(-285,200){$|O_{n_{\cal F},1}|^2$}
\put(-285,200){$|O_{n_{\cal F},1}|^2$}
\put(-195,200){$|O_{n_{\cal F},2}|^2$}
\put(-105,200){$|O_{n_{\cal F},3}|^2$}
\put(-290,15){\scriptsize $n_{\cal F}$}
\put(-200,15){\scriptsize $n_{\cal F}$}
\put(-110,15){\scriptsize $n_{\cal F}$}
\put(-20,15){\scriptsize $n_{\cal F}$}
}
\caption{The Overlaps defined in Eq.(\ref{overlap}) for the lightest four states 
in the $SU(5)$ $k=2$ antisymmetric representation with $P=+$, $q=0$ and flux
tube length $l=32a$.}
\label{Overlap2AL32}
\end{figure}

\begin{figure}[!htb]
\centerline{
\scalebox{1.0}{\input{plot16a} }
\put(-338,245){\scriptsize Ground State}
\put(-250,245){\scriptsize $\rm 1^{st}$ excited state}
\put(-160,245){\scriptsize $\rm 2^{nd}$ excited state}
\put(-70,245){\scriptsize $\rm 3^{rd}$ excited state}
\put(-375,200){$|O_{n_{\cal F},0}|^2$}
\put(-285,200){$|O_{n_{\cal F},1}|^2$}
\put(-285,200){$|O_{n_{\cal F},1}|^2$}
\put(-195,200){$|O_{n_{\cal F},2}|^2$}
\put(-105,200){$|O_{n_{\cal F},3}|^2$}
\put(-290,15){\scriptsize $n_{\cal F}$}
\put(-200,15){\scriptsize $n_{\cal F}$}
\put(-110,15){\scriptsize $n_{\cal F}$}
\put(-20,15){\scriptsize $n_{\cal F}$}
}
\caption{The Overlaps defined in Eq.(\ref{overlap}) for the lightest four states 
in the $SU(5)$ $k=2$ antisymmetric representation with $P=+$, $q=0$ and flux
tube length $l=16a$.}
\label{Overlap2AL16}
\end{figure}

\end{document}